 \documentclass[twocolumn]{emulateapj}

\usepackage{graphicx}
\usepackage{natbib}
\usepackage{txfonts}
\usepackage{color}
\usepackage{graphicx}

\def\lta{\lower2pt\hbox{$\buildrel {\scriptstyle <} 
   \over {\scriptstyle\sim}$}}
\def\gta{\lower2pt\hbox{$\buildrel {\scriptstyle >} 
   \over {\scriptstyle\sim}$}}
\newcommand{\fracb}[2]{\left(\frac{#1}{#2}\right)}

\shorttitle{GRB Dynamics and Afterglow Radiation from AMR-SRHD Simulations}
\shortauthors{F. De Colle et al.}

\begin{document}

\title{Gamma-Ray Burst  Dynamics and Afterglow Radiation from Adaptive Mesh Refinement, Special Relativistic Hydrodynamic Simulations}

\author{Fabio De Colle\altaffilmark{1}, Jonathan Granot\altaffilmark{2,3,4},
Diego L\'opez-C\'amara\altaffilmark{5}, \& Enrico Ramirez-Ruiz\altaffilmark{1}}
\altaffiltext{1}{Astronomy \& Astrophysics Department, University of California, Santa Cruz, CA 95064, USA}
\altaffiltext{2}{Racah Institute of Physics, The Hebrew University, Jerusalem 91904, Israel}
\altaffiltext{3}{Raymond and Beverly Sackler School of Physics \& Astronomy, Tel Aviv University, Tel Aviv 69978, Israel}
\altaffiltext{4}{Centre for Astrophysics Research, University of Hertfordshire, College Lane, Hatfield, AL10 9AB, UK}
\altaffiltext{5}{Instituto de Ciencias Nucleares, Universidad Nacional Aut\'onoma de M\'exico, Ap. 70-543, 04510 D.F., M\'exico}
\email{fabio@ucolick.org}


\begin{abstract}
We report on the development of Mezcal-SRHD, a new adaptive mesh
refinement, special relativistic hydrodynamics (SRHD) code, developed
with the aim of studying the highly relativistic flows in Gamma-Ray
Burst sources. The SRHD equations are solved using finite volume
conservative solvers, with second order interpolation in space and
time. The correct implementation of the algorithms is verified by
one-dimensional (1D) shock tube and multidimensional tests. The code
is then applied to study the propagation of 1D spherical
impulsive blast waves expanding  in a stratified medium with $\rho
\propto r^{-k}$ , bridging between the relativistic and Newtonian
phases (which are described by the Blandford-McKee and Sedov-Taylor
self-similar solutions, respectively), as well as to a two-dimensional
(2D) cylindrically symmetric impulsive jet propagating in a constant
density medium.  It is shown that the deceleration to non-relativistic
speeds in one-dimension occurs on scales significantly
larger than the Sedov length. This transition is further delayed with
respect to the Sedov length as the degree of stratification of the
ambient medium is increased.  This result, together with the scaling
of position, Lorentz factor and the shock velocity as a function of
time and shock radius, is explained here using a simple analytical
model based on energy conservation.  The method used for
calculating the afterglow radiation by post-processing the results of
the simulations is described in detail.  The light curves computed
using the results of 1D numerical simulations during the relativistic
stage correctly reproduce those calculated assuming the self-similar
Blandford-McKee solution for the evolution of the flow. The jet
dynamics from our 2D simulations and the resulting afterglow
lightcurves, including the jet break, 
are in good agreement with those presented in previous works.
Finally, we show how the details of the dynamics critically depend on
properly resolving the structure of the relativistic flow.
\end{abstract}


\keywords{
gamma rays: bursts -  
hydrodynamics - 
methods: numerical - 
relativity
}



\section{Introduction}


Gamma-Ray Bursts (GRBs) are the most electromagnetically
luminous explosions in the Universe.  Their 
non-thermal and highly variable gamma-ray emission implies that the
emitting region must be ultra-relativistic -- moving  with a
very large Lorentz factor, typically $\gtrsim 100$ and sometimes as
high as $\gtrsim 10^3$, in order to avoid excessive pair production at
the source 
\citep[e.g.,][]{2001ApJ...555..540L,Granot08,080916C,090902B,090510-phys}.
At sufficiently large distances from the source the GRB outflow
decelerates as it drives a strong relativistic shock into the
surrounding medium \citep[for reviews see,
e.g.,][]{Piran05,Granot07}. 
Synchrotron emission from this long lived external
shock powers the GRB afterglow, which is observed in the X-rays,
optical or radio, typically over days to months after the prompt
gamma-ray emission. The peak frequency of the afterglow emission
shifts to lower energies as the afterglow shock decelerates by
sweeping up the external medium \citep{1992MNRAS.258P..41R}.
This picture of a decelerating
relativistic expansion of the emitting region during the afterglow
phase is supported by direct measurements of the afterglow image size
at late times in the radio, using very long base-line interferometric
techniques, for GRB~030329 at $z = 0.1685$
\citep{Taylor04,Taylor05,Pihlstrom07}.

GRB activity manifests itself over a dynamical range of
$\sim$ 13 decades in radius \citep{2009ARA&A..47..567G}.
The phenomena  involves different stages, which are usually modeled
separately because of their complexity.  Let us consider these
stages  in turn, working from the small scales to the large scales.

\subsection{Jet Production and the Central Engine}

GRBs divide into two classes according to their duration and
spectral hardness ~\citep{Kouveliotou93}. Long duration GRBs (lasting
$\gtrsim 2\;$s) are associated with Type Ic core collapse SNe, and
thus to the death of massive stars ~\citep{Stanek03,Hjorth03,WB06},
while the nature of short duration GRB (lasting $\lesssim
2\;$s) progenitors is still debated~\citep{LR-R07,Nakar07}, the most
popular model involving the binary merger of two compact stars
~\citep{Paczynski86,Eichler89,Narayan92}. 

In the collapsar model for long GRBs ~\citep{Woosley93}, during the
collapse of a massive Wolf-Rayet progenitor star a black hole is
formed, which rapidly accretes stellar envelope material, launching a
relativistic jet that penetrates the star and eventually powers the
GRB \citep{2002MNRAS.337.1349R}.  It has been modeled using numerical
simulations, where a jet is usually injected as an inner boundary
condition at the center of a collapsing massive star, and bores its
way out of the progenitor star's envelope~\citep{MW99,ZWM03}.  Some
simulations include a magnetic field (in an ideal
magnetohydrodynamical framework) and recently added a general
relativistic
framework~\citep{2004ApJ...606..395M, 2004ApJ...615..389M,
2006ApJ...641..103H, 2006MNRAS.368.1561M, 
Nagataki07,Tchekhovskoy08,Nagataki09,BB11}. The
alternative model for the central engine of long GRBs featuring the
formation of a millisecond magnetar \citep[i.e. a very rapidly
rotating highly magnetized neutron star;][]{Usov92} has also been
studied numerically ~\citep{KB07,Buc07,Buc08,Buc09}. Binary merger
simulations of two neutron stars or a neutron star and a black hole
were performed in the context of short GRBs
~\citep{2002ApJ...577..893L,RR-R02,Rosswog05,Faber06,OJ06,Rezzolla10}.
Recent general relativistic magnetohydrodynamics (MHD) simulations
show that a relativistic jet can naturally form in such a scenario,
which may indeed power short GRBs ~\citep{Rezzolla11}. Similar
simulations of relativistic jet formation from accretion onto a black
hole are routinely performed also in the context of active galactic
nuclei or micro-quasars ~\citep[e.g.,][]{Meier03,KH10}.  Different
processes have been suggested to accelerate and collimate the jet: (i)
thermal energy injected into the jet by annihilation of neutrinos and
anti-neutrinos from an accretion disk
~\citep[e.g.,][]{FW98,Popham99,Ros03,2004ApJ...608L...5L,2006ApJ...641..961L,CB07};
(ii) rotational energy extracted from the central black hole through the
Blandford-Znajek effect ~\citep{BZ77,MR97,BK08}; (iii) rotational
energy extracted from the accretion disk, coupled with a dynamically
important magnetic field ~\citep{BP82,Proga03,LB03,UM06}.

\subsection{Jet Expansion and Deceleration}
Once the GRB outflow transfers most of its energy to the shocked
external medium it becomes dynamically sub-dominant and the flow
becomes insensitive to the exact composition or initial radial
structure of the original outflow. At this stage a spherical flow
approaches the ~\citet[]{BM76} self-similar solution (hereafter BMK),
losing memory of the initial conditions and retaining memory only of
the total energy. The complete evolution of a spherical relativistic
fireball, including the acceleration, coasting and deceleration
phases, has been studied numerically by \citet{KPS99} by using one
dimensional (1D) spherical simulations.

When a non-spherical relativistic outflow (or jet) decelerates, to
zeroth order it locally resembles a section of the spherical BMK
solution characterized by the local value of the energy per solid
angle or isotropic equivalent kinetic energy, $E_{\rm k,iso}$.  Once
the Lorentz factor $\Gamma$ drops to $\theta_0^{-1}$, where $\theta_0$
is the initial half-opening angle of an initially uniform jet with
sharp edges, the jet becomes causally connected in the lateral
direction and can in principal start spreading sideways
significantly. Simple analytic models argue that it should indeed
quickly spread sideways ~\citep{Rhoads97,Rhoads99,SPH99}, while
numerical simulations show that the lateral spreading is much more
modest, and the flow retains memory of $\theta_0$ for a long time,
which for typical values of $\theta_0$ in GRBs lasts up to the
non-relativistic transition time
~\citep{Granot01,GK03,Cannizzo04,ZM09}.

The numerical simulations of jet dynamics during the afterglow stage
are usually done separately from the earlier stages (of the jet
formation, acceleration and collimation), in order to simplify these
challenging numerical computations, which involve a very large
dynamical range.  The most common initial conditions for simulations
of the GRB jet during the afterglow stage are a conical wedge of
half-opening angle $\theta_0$ taken out of the spherical BMK solution
(though in some cases a relativistic cold shell or blob is used
instead). Since the angular size of regions that are casually
connected in the lateral direction is $\sim 1/\Gamma$, such a BMK
wedge should not evolve significantly while its Lorentz factor is
$\Gamma \gg \theta_0^{-1}$, suggesting that the subsequent evolution
should be insensitive to the exact choice of initial Lorentz factor
$\Gamma_0$ in the limit where $\Gamma_0 \gg \theta_0^{-1}$. 

For an ultra-relativistic blast wave most of the energy in the shocked
(downstream) region is within a thin layer behind the shock
transition, whose width is $\Delta\sim 0.1R/\Gamma^2$ in the lab frame
(i.e. the rest frame of the external or upstream medium, which in our
case is also that of the central source), which is hard to resolve
properly for large initial Lorentz factors ~\citep[see,
e.g.,][]{Granot07}. Therefore, most simulations use
$\Gamma_0\theta_0\sim 3-4$ rather than the ideal choice of
$\Gamma_0\theta_0 \gg 1$, along with values of $\theta_0$ that are not
very small (usually $\theta_0 = 0.2$ and $\Gamma_0 \sim 20$), despite
the actual initial Lorentz factors at the onset of the afterglow are
estimated to be at least a few hundred
~\citep[e.g.,][]{2001ApJ...555..540L}, while the values of $\theta_0$
inferred from afterglow observations
~\citep[e.g.,][]{2001ApJ...562L..55F} can be as low as $\sim
0.03-0.05$ (or a high as $\gtrsim 0.5$).

\subsection{Afterglow Jet Simulations: Previous Work and  Goals} 
Since afterglow emission is thought to be predominantly synchrotron
radiation from the shocked external medium, then accurately inferring
the properties of the original relativistic outflow and the external
medium from afterglow observations requires an accurate 
modeling of the dynamics. The jet numerical simulations and
calculations of the corresponding afterglow emission ~\citep{Granot01}
have recently been extended to well within the non-relativistic stage
~\citep[e.g.,][]{ZM09,vanEerten10a,2011arXiv1102.5618W,2011arXiv1105.2485V}. 
Following the dynamics from a
highly ultra-relativistic initial Lorentz factor ($\Gamma_0 > 20$, for
which $\Delta_0/R_0 \sim 10^{-4}(\Gamma_0/30)^{-2}$) down to highly
Newtonian velocities ($v<0.01c$) requires a very large range of
spatial scales, for which an adaptive mesh refinement (AMR) code is
necessary in order to properly calculate the multi-dimensional flow
dynamics.  ~\citet{Granot01} were the first to study this problem
numerically by using multi-dimension numerical simulations and found
that the GRB jet sideways expansion is slower than expected from
analytical models. These results were later confirmed by \citet{ZM09},
who followed the evolution of the GRB jet up to the non-relativistic
phase by running high resolution two-dimensional (2D) simulations.
Simulations using similar initial conditions were also run by
\citet{2010A&A...520L...3M} who found   that
the shock front becomes unstable at high values of the Lorentz factor,
$\Gamma \gtrsim 15$, but the instabilities quickly decay when the jet
decelerate to $\Gamma \lesssim 10$.

All the multi-dimensional numerical simulations of afterglow jets have so far assumed
a uniform external medium, even though a stratified external medium is
expected for the stellar wind of a massive star long GRB progenitor
~\citep{CL00,PK00,2001MNRAS.327..829R,2005ApJ...631..435R}.
This was partly motivated by the faster
deceleration of the afterglow shock with radius in a uniform external
medium compared to a stratified one, which reduces the required
dynamical range of the simulations. Moreover, magnetic fields may also
affect the jet dynamics (in addition to their effect on the afterglow
synchrotron radiation).  ~\citet{2009A&A...494..879M,
2010MNRAS.407.2501M} have used 1D simulations to study the
deceleration of magnetized GRB ejecta propagating into a uniform
ambient medium, and showed that while the late evolution of strongly
magnetized shells resembles that of hydrodynamic shells, the
magnetization plays an important role into the onset of the forward
shock emission. ~\citet{2011arXiv1106.1903M} computed the afterglow
emission produced by a GRB ejecta decelerating into a realistic
external medium by running 1D spherical simulations.
However, multi-dimensional simulations are necessary
in order to fully capture the magnetic field dynamics, as for instance
the generation of turbulence by the magnetohydrodynamics
Kelvin-Helmholtz ~\citep{2009ApJ...692L..40Z} or Richtmyer-Meshkov
~\citep{2008JFM...604..325G} instabilities, and the consequent
magnetic field amplification ~\citep{2010arXiv1011.6350I,
2011ApJ...726...62M}.  Actually, ~\citet{GKS11} have recently shown
that even in 1D one cannot realistically model the deceleration stage
separately from the acceleration stage if the outflow is initially
highly magnetized and accelerates under its own magnetic
pressure. Instead, a full simulation of the acceleration and
deceleration is needed, requiring a very large dynamical range
that is numerically challenging.

With the aim of addressing these questions, and perhaps also possible
applicability to earlier stages of the jet dynamics (such as its
acceleration or propagation within the progenitor star), we have
developed a new AMR, relativistic hydrodynamic code.  While the code
developed is similar in several aspects to previous SRHD-AMR codes
~\citep[e.g.,][]{2002ApJ...572..713H, 2005ApJ...635..723A, 
ZM09, 2007MNRAS.376.1189M, 2007ApJ...665..569M, Wang08}
we consider it important to present a detailed,
self-contained description of the hydrodynamic code as well as the
matching radiation code, along with detailed tests.
The paper is organized as follows.  
\S \ref{sec2} and \S \ref{sec3} describe in
detail, respectively, the SRHD code and the radiation code used to
calculate the observed afterglow emission (by post-processing the
outcome of the SRHD simulation). Standard tests used to verify the
SRHD code are presented in the Appendix, while the correct
implementation of the radiation code is discussed in \S \ref{sec4}. 
\S \ref{sec4} presents a detailed study of the propagation of a relativistic, purely hydrodynamic
ejecta into a one-dimensional stratified medium as well in a multi-dimensional
homogeneous medium together with the resulting lightcurves.  Finally, \S \ref{sec5}
presents our conclusions.  Simulations of the propagation of jets
into a stratified medium and the inclusion of magnetized flows will be
addressed in future work.

\section{Numerical code}
\label{sec2}

\subsection{Relativistic Hydrodynamics equations}

The special relativistic hydrodynamics (SRHD) equations in
conservative form ~\citep[e.g.,][]{1989cup..book.....A} can be
written as follows:
\begin{eqnarray}
 {\partial D \over\partial t}
  +\nabla\cdot\left(\,D \vec v \,\right)  = 0  
 \label{rhd1:eq}
  \\
 {\partial \vec S \over\partial t}
  +\nabla\cdot
   \left( \vec S \vec v  + p \mathcal{I} \right)= 0
 \label{rhd2:eq}
 \\
 {\partial \tau \over\partial t}
  +\nabla\cdot
  \left( \tau \vec v+ p\vec v \right) = 0
 \label{rhd3:eq}
\end{eqnarray}
where $p$ is the thermal pressure, $\vec{v} = \vec{\beta}c$ is the
flow velocity ($c$ being the speed of light), and $\mathcal{I}$ is the
identity matrix.  These equations represent the conservation of rest
mass (\ref{rhd1:eq}), momentum (\ref{rhd2:eq}), and energy
(\ref{rhd3:eq}).  The conserved variables ($D, \vec S, \tau$) correspond
to the lab frame rest mass, momentum, and energy (excluding rest
mass) densities, respectively. They are related to the primitive
variables ($\rho, \vec v, p$) by the following relations:
\begin{eqnarray}
  D = \rho \Gamma \;,
  \label{eq:D}
  \\
  \vec S = D h \Gamma \vec v \;,
  \label{eq:s}
  \\
  \tau = D h \Gamma c^2 - p - D c^2 \;,
  \label{eq:tau}
\end{eqnarray}
where $\Gamma = (1-\beta^2)^{-1/2}$ is the Lorentz factor, $\rho$ is
the proper rest mass density, and $h$ is the specific enthalpy.  The
SRHD system of equations is closed by the equation of state, relating
$h$ to $p$ and $\rho$.  Note that by explicitly subtracting the
rest mass in the definition of the lab frame energy density $\tau$ in
equation~(\ref{eq:tau}), the non-relativistic hydrodynamic equations
are properly recovered when $\beta\ll 1$.

\subsection{Integration methods}
\label{sec:int}

The SRHD equations~(\ref{rhd1:eq})-(\ref{rhd3:eq}) form an hyperbolic
system of equations and can be  solved by using methods similar
to those developed for classical non-relativistic gas dynamics
\citep[for a review see, e.g.,][]{Toro08}.  Without loss of generality,
the solution of the hyperbolic system of equations
\begin{equation}
  {\partial u \over \partial t} + \nabla \cdot \vec f = 0\ ,
\end{equation}
is given in 1D (the generalization to multi-dimensions is
straightforward) by:
\begin{equation}
  U_{i}^{n+1} = U_{i}^{n} - \frac{\Delta t}{\Delta x_i}
    (F_{i+1/2}^{n+1/2}-F_{i-1/2}^{n+1/2})\ ,
  \label{eq:cons}
\end{equation}
where $x_i$ represent the position of the center 
of the cell $i$ with volume $\Delta x_i=x_{i+1/2}-x_{i-1/2}$,
$x_{i\pm1/2}$ are the positions of the interfaces between 
the cells $x_i$ and $x_{i\pm 1}$, and
\begin{equation}
  U_{i}^n = \frac{1}{\Delta x_i} \int_{x_{i-1/2}}^{x_{i+1/2}} u_i(t^n,x) dx \;,
\end{equation}
\begin{equation}
  F_{i\pm1/2}^{n+1/2} = \frac{1}{\Delta t} \int_{t_n}^{t_{n+1}} 
                f(t,x_{i\pm1/2}) dt\ ,
  \label{eq:flux}
\end{equation}
are the volume average of the conservative variables and their time-averaged fluxes.

While equation~(\ref{eq:cons}) represents an exact solution of the
corresponding partial differential equation, an approximation is
introduced when the fluxes (equation~\ref{eq:flux}) are computed.
Because an exact solver is in general very expensive, in the current
version of the code we have implemented the simple and computationally
efficient relativistic extension \citep{1993JCoPh.105...92S} of the HLL
method \citep{1983JCoPh..49..357H}.

It is well-known that the HLL method does not resolve properly the 
contact discontinuity, and it has an intrinsic high level of numerical 
diffusivity, while for instance other methods (e.g. the HLLC method,
 \citealt{Mignone05}) properly reconstructs the
contact discontinuity, producing results with significantly  lower dissipation. 
On the other hand, 
being more diffusive, the HLL method is also more
``robust'', very rarely producing  unphysical results such as negative
pressures or imaginary Lorentz factors.  In addition, a low
dissipation method may produce undesirable effects, such as a
``carbuncle'' artifact along the axis of propagation of strong shocks
\citep[see the discussion by][]{Wang08}.

Second order accuracy in time and space are obtained 
by employing a Runge-Kutta integrator and by a spatial 
reconstruction of the primitive variables \citep{1979JCoPh..32..101V}, 
except in shocks where the methods drops to first order 
(in space) by a limiter. 
Different limiters are implemented, including the
``minmod'' (being the most diffusive), UMIST, Superbee and 
the less diffusive ``monotonized central 
difference'' limiter.

\subsection{Extension to cylindrical and spherical coordinates}

The extension to cylindrical and spherical coordinates is treated very carefully 
in the code.
For instance, in two-dimensional ($r$, $\theta$) spherical coordinates, the
equations read:
\begin{equation}
  \frac{\partial U}{\partial t} + \frac{1}{r^2} \frac{\partial(r^2 F)}{\partial r} +
        \frac{1}{r\sin\theta}
        \frac{\partial(G\sin\theta)}{\partial\theta} = \frac{S}{r}\ ,
  \label{eq:sph}
\end{equation}
where $U$, $F$, $G$, $S$ can be easily derived  from 
equations~(\ref{rhd1:eq})-(\ref{rhd3:eq}). 
We note that a simple cell-center discretization of this system of
equations introduces large numerical errors when differencing. In
particular, it does not preserve stationary initial conditions to
machine accuracy. As an example, if one assumes static initial conditions, as $\partial p / \partial r = 0$, $\rho$ constant 
and $\vec v=0$, these  are preserved in the code if, 
e.g., the relation (easily derived from the $\theta-$component of 
the momentum equation)
\begin{equation}
\frac{1}{r \sin \theta} \frac{\partial(p\sin\theta)}{\partial \theta} = 
\frac{p}{r}\frac{\cos \theta}{\sin \theta}\ ,
  \label{eq:he}
\end{equation}
is held to machine accuracy. A simple centered discretization gives
\begin{equation}
\frac{1}{\sin \theta_{j}} \frac{\sin \theta_{j+1/2}
-\sin \theta_{j-1/2}}{\theta_{j+1/2}-\theta_{j-1/2}} 
\neq \frac{\cos \theta_{j}}{\sin \theta_{j}}\ ,
\end{equation}
where $\theta_j$ is evaluated at the center of the cell, while $\theta_{j\pm1/2}$
at the interface between different cells, and it does not preserve the initial
conditions.

A way to minimize numerical errors when differencing equation~(\ref{eq:sph}), 
especially near coordinate singularities, is by a finite volume discretization 
\citep[e.g.,][]{1991MNRAS.250..581F, li03}, that is by averaging the variables over the cell volume. 
Given for instance the cell centered in $(i,j)$ and with nodes located 
at $(i\pm1/2,j\pm1/2)$, the value of of the quantity $A$ averaged over the cell volume is given by
\begin{equation}
\langle A\rangle = \frac{\int  \sin \theta d\theta \int A r^2 dr}{\int  \sin \theta d\theta \int r^2 dr}\ .
\end{equation}

With this definition, radial and polar derivatives
are approximated by (taking $A = \frac{1}{r^2} \frac{\partial(r^2 F)}{\partial r}$
and $A =   \frac{1}{r \sin \theta} \frac{\partial(G\sin \theta )}{\partial \theta}$ respectively):
\begin{eqnarray}
  \frac{1}{r^2} \frac{\partial(r^2 F)}{\partial r} \approx 
  \frac{\delta_i(r^2 F)}{\delta_i(r^3/3)}  \;, \nonumber \\
  \frac{1}{r \sin \theta} \frac{\partial(G\sin \theta )}{\partial \theta} \approx 
  \frac{\delta(G \sin \theta)}{-\delta(\cos \theta)}
  \frac{\delta(r^2/2)}{\delta(r^3/3)} \;,
\end{eqnarray}
where $\delta_i(f) = f_{i+1/2}-f_{i-1/2}$, while the source terms are
discretized by assuming (taking $A=\frac{1}{r}$ and $A=\frac{\cos \theta}{\sin \theta}$ respectively):
\begin{equation}
  \frac{1}{r} \approx 
  \frac{\delta_i(r^2/2)}{\delta_i(r^3/3)}  \;, \qquad
  \frac{\cos \theta}{\sin \theta} \approx 
  \frac{\delta(\sin \theta)}{-\delta(\cos \theta)} \;.
\end{equation}

It is easy to verify that, written in this form, 
equation~(\ref{eq:he}) preserves static initial conditions to 
machine accuracy.

\subsection{Equation of state}

The equation of state relates the enthalpy to the pressure and
density.  In the case of a relativistic perfect gas it takes the form
\citep{1971tar..book.....S}
\begin{equation}
  h = \frac{K_3(1/\Theta)}{K_2(1/\Theta)}\ ,
\label{eos:synge}
\end{equation}
where $\Theta=p/(\rho c^2)$, and $K_i$ is the $i^{\rm th}$-order of
the modified Bessel functions of the second kind.

As the evaluation of the enthalpy from equation~(\ref{eos:synge}) is 
computationally expensive (see e.g. \citealt{1996MNRAS.278..586F}), simplified relations 
have been used, the simplest being the $\bar\gamma$-law equation of state (EOS) 
\begin{equation}
  h = 1 + \frac{\bar\gamma}{\bar\gamma-1} \Theta\ ,
\label{eos:gamma}
\end{equation}
with a constant value of the adiabatic index $\bar\gamma$ fixed and
equal to $4/3$ or $5/3$, valid only in the limit of ultra-relativistic
or sub-relativistic fluids, respectively.

\citet[]{Mignone05} proposed the EOS (see also \citealt{Mathews71})
\begin{equation}
  h = \frac{5}{2}\Theta + \frac{3}{2}\sqrt{\Theta^2+\frac{4}{9}}\ ,
\label{eos:mathews}
\end{equation}
which in addition to approximating equation~(\ref{eos:synge}) 
within 2\%, also satisfies the \citet{Taub48} inequality
\begin{equation}
  (h-\Theta)(h-4\Theta) \leq 1\ ,
\label{eos:taub}
\end{equation}
in accordance with relativistic kinetic theory.

More recently, \citet{Ryu06} proposed a simpler and better
approximation to the Synge EOS (accurate to within 0.5\%), which also satisfies
the Taub inequality (equation~\ref{eos:taub}), given by
\begin{equation}
  h = 2\frac{6 \Theta^2 + 4 \Theta + 1}{3 \Theta + 2}
\label{eos:ryu}
\end{equation}
The implementation of these EOS is straightforward, and unless stated
otherwise, in this paper we use the one derived by \citet{Ryu06}.

\subsection{Converting conserved to primitive variables}

The increased level of complexity in solving the SRHD equations when
compared to the corresponding non-relativistic hydrodynamics equations arises
mainly from the lack of simple closed expressions relating conserved
($\tau$, $\vec S$, $D$) and primitive ($p$, $\vec v$, $\rho$)
variables. This requires the primitive variables to be computed from
the conserved variables by a non-linear iteration.

Among others, \citet{2006ApJ...641..626N} studied several algorithms to convert 
conserved to primitive variables for the case of a $\bar\gamma$-law EOS.
\citet{Ryu06}, for the EOS defined in equation~(\ref{eos:ryu}), applied a 
Newton-Raphson method to an 8-th order equation dependent on 
$\Gamma$. \citet{2007MNRAS.378.1118M}, for the 
case of relativistic MHD with a general equation of state, 
derived an equation for $W=D\rho h$, and evaluate $W$ 
by a Newton-Raphson iterative scheme, with
the derivative $dW/dp$ given by using thermodynamics relations.
Here, we present a different implementation.
Taking advantage of the existence of a relation between
the specific enthalpy $h$ and $\Theta=p/(\rho c^2)$,
we solve the system of equations (\ref{eq:D}-\ref{eq:tau})
as a function of $\Theta$
by using a standard Newton-Raphson method, 
and  we then determine the other variables.

First, squaring the momentum equation ($S_k=Dh\Gamma v_k$) we get:
\begin{equation}
 \Gamma^2 = 1 + \frac{S^2}{D^2 h^2}
 \label{eq:gam}
\end{equation}
with $h=h(\Theta)$.  From the definition of specific enthalpy it
 follows that $h \geq 1$. Therefore, equation~(\ref{eq:gam}) 
leads to the following inequality (e.g. Schneider et al. 1993)
\begin{equation}
 1 \le \Gamma^2 \leq 1 + \frac{S^2}{D^2}\ .
 \label{eq:cond}
\end{equation}

By using the relation $p = D \Theta c^2/\Gamma$, we can then derive from the
definition of energy density (excluding rest mass, i.e. $\tau = D h \Gamma
c^2 - p - D c^2$) the following identity
\begin{equation}
 f(\Theta) = h(\Theta) \Gamma(\Theta) - \frac{\Theta}{\Gamma(\Theta)} 
- 1 - \frac{\tau}{D c^2}= 0\ .
 \label{eq:fth}
\end{equation}
Equations~(\ref{eq:gam}) and (\ref{eq:fth}) are then used, 
together with a standard Newton-Raphson method, to determine 
$\Theta$, with $df/d\Theta$ given by
\begin{equation}
\frac{df(\Theta)}{d\Theta} =  \frac{h^\prime}{\Gamma}  
\left(1 - \frac{\Theta}{h}
   \frac{\Gamma^2-1}{\Gamma^2} \right)
 - \frac{1}{\Gamma}\ ,
\end{equation}
where the relation
$\Gamma^\prime = - h^\prime (\Gamma^2-1)/(h \Gamma)$
has been used (derived from equation~\ref{eq:gam}), and $h^\prime = dh/d\Theta$. 
The derivative $dh/d\Theta$ depends on the particular
EOS used, and can be determined both analytically or numerically.
In the case of the \citet{Ryu06} EOS (equation~\ref{eos:ryu}),
$h^\prime =  4 - 6/(3\Theta+2)^2$.

We also note that $df(\Theta)/d\Theta > 0$ for every value of $\Theta$ (for the 
EOS considered here). Therefore, as $f(\Theta\to\infty)>0$,
a solution for the equation $f(\Theta)=0$ exists if
$f(\Theta=0)<0$, which implies the relation 
\begin{equation}
  D^2+S^2 < (D+\tau/c^2)^2\ ,
\end{equation}
must hold in order to allow a solution with physically 
acceptable values of $\Gamma$ and $p$ (that is, real values
of $\Gamma \geq 1$ and $p \geq 0$).

As we have shown, this method can be easily applied to any 
equation of state of the form $h=h(\Theta)$.
Furthermore, the guess used by  the Newton-Raphson method (NRM) is provided
by simply assuming $\Theta=0$. 
In this case, setting a tolerance of 10$^{-10}$ into
the Newton-Raphson solver, the method converges typically 
within $\sim 5$ iterations.
In very rare cases when the
NRM fails to converge, a bisection method is used instead.

\subsection{Adaptive mesh refinement}

We have implemented the SRHD equations in the framework of the
adaptive mesh refinement code Mezcal.
In the code, a basic Cartesian grid is built at the beginning of
the simulation, and it is refined based on the initial conditions and
the subsequent  evolution of the flow. The uniform version of
the code has been used in the past to simulate MHD jets 
\citep[e.g.,][]{2005MNRAS.359..164D, 2006A&A...449.1061D,
2008ApJ...689..302D}.

In the Mezcal code, the computational grid is divided in ``octs'' (or
blocks) of 2$^{n_{\rm dim}}$ cells, where $n_{\rm dim}$ is the number of dimensions 
of the problem. Each block has a series of pointers to its vertexes,
and each vertex has pointers to the octs sharing that particular
vertex. In this way, neighbor octs (both along the axes and the
diagonal direction) can be easily located in the grid, facilitating
the computation of the MHD solver (that, in staggered mesh methods, 
is based on determining electric fields at the cell vertexes).  
At a given time, each position on the grid is covered by only one cell,
i.e. there are no pointers between ``parent'' and ``sibling'' usually present
in other tree-AMR codes \citep[e.g.][]{1984JCoPh..53..484B,1998JCoPh.143..519K}.
Furthermore, there are no ghost
cells in any of the blocks. 
Although the use of pointers causes a small memory overload 
(corresponding to 4 integers per cell in 3 dimensions), that is largely 
compensated by the fact that, due to the small block size, the grid covers 
only regions that effectively need to be refined.

At every timestep, all blocks are swept, and they are
refined/coarsened if a user defined criterion is fulfilled. Typically,
this criterion is based on the first or second derivative of some variable, but more complex criteria can be easily implemented.
Once a list of blocks flagged for refinement has been formed, the grid
is checked for consistency. As the code maintains a maximum ratio of 2
in the size of neighbor cells, all coarser neighbors of blocks are flagged for refinement.  When a block is
refined, 2$^{n_{\rm dim}}$ new blocks are created, and the parent block is
eliminated.  To avoid excessive memory fragmentation, the block lists
are periodically reordered.

Coarsening is allowed only when the 2$^{n_{\rm dim}}$ neighbor blocks (previously produced 
by refining the same parent block) are marked for derefinement 
during the same timestep.
We use zeroth-order interpolation when refining, and we integrate
the conserved variables over the volume when coarsening, following
the  strategy presented by \citet{li03}.

To evolve the hyperbolic equations, the code employs a timestep common
to all grid levels.
While the use of a global timestep may potentially produce an important 
computational overload \citep[as large of 50\%, depending on the problem, see e.g.][]
{2003astro.ph.10891D} with respect to using a local timestep, the local time step 
method can represent an important bottle-neck for parallelization, 
as blocks on different levels must to be evolved sequentially (and not in parallel).
The fluxes are computed by locating the neighbor blocks, and considering
the cells sharing the same faces.
When two blocks with different levels of refinement share the same 
face, $2^{n_{\rm dim}-1}$ fluxes are computed between the 2$^{n_{\rm dim}-1}$ cells located on the 
higher level block and the cell part of the block at the lower level
of refinement. The fluxes are then added to the conserved variables of the 
cells sharing the common boundary.

The Mezcal code is parallelized by using the \emph{Message Passing Interphase} (MPI) 
library. 
The communication time is minimized by scheduling it in parallel
with the calculation of the fluxes. This is done by first computing
the fluxes between blocks located in each process, and then, 
once the communication phase is completed, computing the 
rest of the fluxes (between blocks ``inside'' each process
and ghost blocks).
The load balancing is achieved by ordering the blocks by
a space-filling curve \citep{sagan94}, dividing the total number of blocks 
between the different processes, and moving blocks between unbalanced 
processes. In the code, the Morton and the Hilbert space-filling curves
\citep{sagan94} are implemented. The load balancing is typically 
applied every $\sim$10 timestep, and represents an overload of $\sim$1\% of the 
total computational time. The parallel scaling of the AMR code is under
evaluation and will be presented elsewhere.

\section{Calculation of the emitted 
radiation  from a hydrodynamic simulation}
\label{sec3}

\subsection{Calculation of the observed flux density}

Here we provide a detailed derivation  of the procedure required to calculate 
the radiation emitted from a relativistic source, following
\citet{GR-R11}, which is based on previous work
\citep{GPS99a,GPS99b,GK03,KG03}. 


\begin{figure}
\centering
 \includegraphics[width=0.45\textwidth]{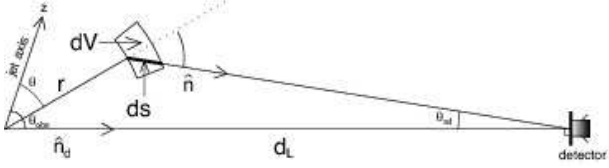}
\caption{The contribution of a volume element $dV$ to the flux observed by
a distant observer is $dF_\nu(\hat{n}_{\rm d}) =
I_\nu(\hat{n})\cos\theta_{\rm sd}\,d\Omega_{\rm sd} \approx
I_\nu(\hat{n})\,d\Omega_{\rm sd}$, where $\theta_{\rm sd}$ is the
angle between the direction opposite to that at which the detector is
pointing ($\hat{n}_d = \hat{Z}$ in the figure) and the local direction
from a small emitting region within the source (of volume $dV$) to the
detector. Since the observer is far away, the direction of emission in
the observer frame is almost parallel to the z-axis.}
\label{fig1}
\vskip 0.5cm
\end{figure}

The geometry of the problem is shown in Figure~\ref{fig1}.  
We denote with $\theta_{\rm sd}$ the angle subtended by the
direction $\hat{n}_{\rm d}$ of the observer (perpendicular to the
differential area $dA$ at the detector, and opposite to the direction
at which the detector is pointing) and the local direction $\hat{n}$ from the relevant
(contributing) part of the source to the observer. In practice
almost always $\theta_{\rm sd}\ll 1$, as the source size is much
smaller than the distance from the source to the observer, so that
$\cos\theta_{\rm sd} \approx 1$.  We also define $d\Omega_{\rm sd} =
d\phi_{\rm sd}d\cos\theta_{\rm sd}$ as the differential solid angle
subtended by the contributing portion of the source as viewed by the
observer.  Our aim is to calculate the observed flux density, $F_\nu =
dE/dAd\nu dt$, which is the energy per unit area, frequency and time
in the direction $\hat{n}_d$ normal to $dA$.  From the definition of
the angular distance to the source, $d_A(z)$, where $z$ is the
cosmological redshift, we have $d\Omega_{\rm sd} = dS_\perp/d_A^2$,
where $dS_\perp$ is the differential area in the plane of the sky
(normal to $\hat{n}$) sustained by the source.  The angular distance
to the source is related to the luminosity distance: $d_L(z)$, by $d_A
= (1+z)^{-2}d_L$.

The differential contribution to the flux can be written as
$dF_\nu(\hat{n}_{\rm d}) = I_\nu(\hat{n})\cos\theta_{\rm
sd}\,d\Omega_{\rm sd} \approx I_\nu(\hat{n})\,d\Omega_{\rm sd} = I_\nu
dS_\perp/d_A^2$.  Here $I_\nu(\hat{n}) = dE/dAd\Omega d\nu dt$ is the
specific intensity (the energy per unit area, time and frequency of
radiation directed within a small solid angle $d\Omega$, which is
centered on the direction $\hat{n}$), and should be evaluated at the
location of the observer.

For an optically thin source
$I_{\nu_z} = \int j_{\nu_z} ds_z$, where $j_{\nu_z} =
dE_z/dV_zd\Omega_z d\nu_z dt_z$ is the emitted energy per unit volume,
solid angle, frequency and time, while $ds_z$ is the differential path
length along the trajectory of a photon that reaches the observer at
the time $t_{\rm obs}$ when $F_\nu$ is measured  (the subscript $z$ here denotes quantities measured in the
cosmological frame of the source).   
Since $I_\nu/\nu^3$,
$j_\nu/\nu^2$ and $ds/\nu$ are Lorentz invariant \citep{RL79}, we have
$I_\nu = (\nu/\nu_z)^3I_{\nu_z} = (1+z)^{-3}\int j_{\nu_z}ds_z$. 
Therefore, $dF_\nu(\hat{n}_{\rm d}) = I_\nu dS_\perp/d_A^2 =
j_{\nu_z}dV_z\,(1+z)/d_L^2$, where $dV_z = dS_\perp ds_z$ is the
volume element in the source cosmological frame. Here $j_{\nu_z} =
[\Gamma(1-\hat{n}\cdot\vec{\beta})]^{-2}\,j'_{\nu'}$ is measured in
the source (cosmological) frame, while $j'_{\nu'}$ is measured in the
(comoving) rest frame of the emitting material, which expands at a
velocity $\vec{\beta}c$ in the source frame. Altogether, this
gives\footnote{Since $\theta_{\rm sd}\ll 1$ we make the
approximation that $\hat{n}_d \approx \hat{n}$, and replace
$F_\nu(t_{\rm obs},\hat{n}_d)$ with $F_\nu(t_{\rm obs},\hat{n})$, for
simplicity. As long as $\theta_{\rm sd}\ll \Gamma^{-1}$ (which is also
typically the case, in particular for cosmological GRBs) one can also
neglect the change in $\hat{n}$ between the different parts of the
source, and replace $\hat{n}$ with the constant $\hat{n}_d$.}
\begin{equation}\label{F_nu1}
F_\nu(t_{\rm obs},\hat{n}) = \frac{(1+z)}{d_L^2(z)}\int d^4x\;\delta
\left(t_z-\frac{\hat{n}\cdot\vec{r}}{c}-\frac{t_{\rm obs}}{1+z}\right)
\frac{j'_{\nu'}}{\Gamma^2(1-\hat{n}\cdot\vec{\beta})^2}\ ,
\end{equation}
where $t_z$ is the coordinate time at the source's cosmological frame,
\begin{equation}
\nu' = (1+z)\Gamma(1-\hat{n}\cdot\vec{\beta})\nu\ ,\quad\quad
t_{\rm obs} = (1+z)\left(t_z-\frac{\hat{n}\cdot\vec{r}}{c}\right)\ ,
\end{equation}
and $t_{\rm obs} = 0$ corresponds to a photon emitted at the origin
($\vec{r}=0$) at $t_z=0$.  Since $d^4x = dt_zdV_z = dt_zdS_\perp ds_z
= dt_zdS_\perp ds'(\nu_z/\nu') =
dt_zdV'/\Gamma(1-\hat{n}\cdot\vec{\beta})$ and $4\pi j'_{\nu'}dV' =
dL'_{\nu'} = 4\pi(dE'/d\Omega'd\nu'dt')$ is the differential of the
isotropic equivalent spectral luminosity in the comoving frame,
equation~(\ref{F_nu1}) can be rewritten as
\begin{eqnarray}\label{F_nu2}
F_\nu(t_{\rm obs},\hat{n}) = \frac{(1+z)}{4\pi d_L^2(z)}&\int&
dt_z\;\delta \left(t_z-\frac{\hat{n}\cdot\vec{r}}{c}-\frac{t_{\rm obs}}{1+z}\right) \times\nonumber \\
\times &\int& \frac{dL'_{\nu'}}{\Gamma^3(1-\hat{n}\cdot\vec{\beta})^3}\ .
\end{eqnarray}

There are two main approaches to calculate $F_\nu$ from the results of
a numerical simulation. The first one relies on numerically
calculating $I_\nu$ along different lines of sight
(i.e. trajectories or world lines of photons that reach the
observer), and then computing $dF_\nu = I_\nu dS_\perp/d_A^2$. This
was applied both in analytical \citep{GPS99b,GS02} and in numerical
~\citep{Salmonson06,vanEerten10a,vanEerten10b} calculations. Its main
advantages are that it can properly handle the optically thick regime,
where the radiative transfer equation is solved (analytically or
numerically) along each line of sight, and that it provides the
observed image of the source (i.e. $I_\nu$ on the plane of the sky) as
a by-product, since it is used when calculating $F_\nu$. Its main
disadvantage for numerical simulations is that it requires accessing
many different ``snapshots'' of the simulation results, corresponding
to different lab frame times $t_z$, for calculating each value of
$I_\nu$, as it requires integration along the trajectories (or
world-lines) of photons that reach the observer. The second approach,
we adopt here, avoids this difficulty, and was already used in several
previous studies
\citep{Granot01,Granot02,GK03,KG03,NG07,ZM09}.  In this approach the
range of observed times, $t_{\rm obs}$, is divided into a finite
number ($N_t$) of time bins of width $\Delta t_{{\rm obs,}i}$ centered
on $t_{{\rm obs,}i}$ (for $i = 1,...,N_t$). That is, the $i^{\rm th}$
bin corresponds to $t_{{\rm obs,}i}-\Delta t_{{\rm obs,}i}/2 < t_{\rm
obs} < t_{{\rm obs,}i}+\Delta t_{{\rm obs,}i}/2$, and there are no
overlaps or gaps, so that $t_{{\rm obs,}i}+\Delta t_{{\rm
obs,}i}/2=t_{{\rm obs,}i+1}-\Delta t_{{\rm obs,}i+1}/2$ for $1\leq
i\leq N_t-1$. For many physical systems (such as the ones we simulate)
it is convenient to choose logarithmically spaced bins, with a
constant $\Delta t_{{\rm obs,}i}/t_{{\rm obs,}i}$. If the time bins
are sufficiently densely spaced, such that the second time derivative
(with respect to $t_{\rm obs}$) of $F_\nu$ is correspondingly small,
then $F_\nu(t_{{\rm obs,}i},\hat{n})$ can be approximated by its
average value within the $i^{\rm th}$ time bin,
\begin{equation}
F_\nu(t_{{\rm obs,}i},\hat{n}) = \frac{1}{\Delta t_{{\rm obs,}i}}
\int_{\Delta t_{{\rm obs,}i}-\Delta t_{{\rm obs,}i}/2}
^{\Delta t_{{\rm obs,}i}+\Delta t_{{\rm obs,}i}/2}
dt_{\rm obs}F_\nu(t_{\rm obs},\hat{n})\ .
\end{equation}
Now given that $\delta[f(x-x_0)] = \delta(x-x_0)/|f'(x_0)|$ when $f(x)$
has a single root at $x_0$, we obtain
\begin{widetext}
\begin{eqnarray}\nonumber
F_\nu(t_{{\rm obs,}i},\hat{n}) &=& 
\frac{(1+z)}{d_L^2(z)\Delta t_{{\rm obs,}i}}\int d^4x
\int_{\Delta t_{{\rm obs,}i}-\Delta t_{{\rm obs,}i}/2}
^{\Delta t_{{\rm obs,}i}+\Delta t_{{\rm obs,}i}/2}dt_{\rm obs}
\;\delta\left(\frac{t_{\rm obs}}{1+z}-t_z+\frac{\hat{n}\cdot\vec{r}}{c}\right)
\frac{j'_{\nu'}}{\Gamma^2(1-\hat{n}\cdot\vec{\beta})^2}
\\ \label{eq:F_nu}
 &=& \frac{(1+z)^2}{d_L^2(z)\Delta t_{{\rm obs,}i}}\int d^4x\;
H\left(\frac{\Delta t_{{\rm obs,}i}}{2(1+z)}
-\left|\frac{t_{{\rm obs,}i}}{1+z}-t_z+\frac{\hat{n}\cdot\vec{r}}{c}\right|\right)
\frac{j'_{\nu'}}{\Gamma^2(1-\hat{n}\cdot\vec{\beta})^2}
\\ \nonumber
 &=& \frac{(1+z)}{d_L^2(z)\Delta t_{{\rm obs,}z,i}}\int d^4x\;
H\left(\frac{\Delta t_{{\rm obs,}z,i}}{2}
-\left|t_{{\rm obs,}z,i}-t_z+\frac{\hat{n}\cdot\vec{r}}{c}\right|\right)
\frac{j'_{\nu'}}{\Gamma^2(1-\hat{n}\cdot\vec{\beta})^2}\ ,
\end{eqnarray}
\end{widetext}
where $H(x)$ is the Heaviside step function and $t_{{\rm obs},z}\equiv t_{\rm obs}/(1+z)$.

The results of a simulation that models the dynamics of a
physical system are naturally given at a finite number ($n_t$) of time
steps ($t_{z,j}$, where $j = 1,...,n_t$), i.e. ``snapshots'' of the
dynamics. At each snapshot the values of the hydrodynamic variables
are provided at a finite number of points, each at the center of a
computational cell, which represents a finite three dimensional volume
$\Delta V^{(3)}$ (generally different from that of other cells for an
AMR code). For this reason, we assign to each snapshot time $t_{z,j}$ a finite time
interval: $(3t_{z,1}-t_{z,2})/2 < t_z < (t_{z,1}+t_{z,2})/2$ and
$\Delta t_{z,1} = t_{z,2}-t_{z,1}$ for $j=1$, $(t_{z,j-1}+t_{z,j})/2 <
t_z < (t_{z,j}+t_{z,j+1})/2$ and $\Delta t_{z,j} =
(t_{z,j+1}-t_{z,j-1})/2$ for $2\leq j\leq n_t-1$, and
$(t_{z,n_t-1}+t_{z,n_t})/2 < t_z < (3t_{z,n_t}-t_{z,n_t-1})/2$ and
$\Delta t_{z,n_t} = t_{z,n_t}-t_{z,n_t-1}$ for $j=n_t$. Sufficiently
dense and well distributed snapshot times are key to the flux calculations. Thus, the simulation
provides a finite number of 4-dimensional space-time cells, which
together cover the finite simulated 4-volume (the time and three
dimensional volume covered by the simulation\footnote{Note that
usually even in a 1D or 2D simulation represents a 3D volume given the
relevant assumed symmetry of the problem.}).  The 4-volume of the
$k^{\rm th}$ 3D cell of the $j^{\rm th}$ snapshot time is $\Delta
V^{(4)}_{jk} = \Delta t_{z,j}\Delta V^{(3)}_{jk}$. Given the physical
conditions in each such 4D space-time cell we can then calculate its local
(comoving) emissivity, $j'_{\nu'}$ (under appropriate assumptions) and
use equation~(\ref{eq:F_nu}) in order to calculate its contribution to the
observed flux density, $F_\nu$. 
The proper way of doing this is 
to calculate the fraction $f_{ijk}$ of its 4-volume $\Delta
V^{(4)}_{jk}$ that falls within each observer time bin centered on
$t_{{\rm obs,}i}$, resulting in the following discretized version of
equation~(\ref{eq:F_nu}),
\begin{equation}\label{eq:F_nu-dis}
F_\nu(t_{{\rm obs,}i},\hat{n}) = \frac{(1+z)^2}{d_L^2(z)\Delta t_{{\rm obs,}i}}
\sum_{j,k}f_{ijk}\,\Delta V^{(4)}_{jk}
\frac{j'_{\nu',jk}}{\Gamma^2_{jk}(1-\hat{n}\cdot\vec{\beta}_{jk})^2}\ ,
\end{equation}
where the subscript ``$jk$'' indicates that the relevant quantities
are evaluated at the appropriate cell, centered on $(t_z,\vec{r}) =
(t_{z,j},\vec{r}_{jk})$. Since the order of the summation is not
important, it is much more convenient to evaluate the contributions of
each 4D cell according to the order at which it is stored.  Since it is not
always convenient and may cost additional computational time to
calculate all of the coefficients $f_{ijk}$, one might further
simplify equation~(\ref{eq:F_nu-dis}) by attributing all of the
contribution from any given 4D cell to a single observer time
interval, corresponding to that of the cell's center:
\begin{widetext}
\begin{eqnarray}\nonumber
\Delta F_{\nu,i,jk}(\hat{n}) = \frac{(1+z)^2}{d_L^2(z)}
\frac{\Delta V^{(4)}_{jk}}{\Delta t_{{\rm obs,}i}}
\frac{j'_{\nu',jk}}{\Gamma^2_{jk}(1-\hat{n}\cdot\vec{\beta}_{jk})^2}
\quad{\rm for}\quad 
\left|\frac{t_{{\rm obs,}i}}{1+z}-t_{z,j}+\frac{\hat{n}\cdot\vec{r}_{jk}}{c}\right|
<\frac{\Delta t_{{\rm obs,}i}}{2(1+z)}
\\ \label{eq:F_nu-dis1}
= \frac{(1+z)}{d_L^2(z)}
\frac{\Delta V^{(4)}_{jk}}{\Delta t_{{\rm obs,}z,i}}
\frac{j'_{\nu',jk}}{\Gamma^2_{jk}(1-\hat{n}\cdot\vec{\beta}_{jk})^2}
\quad{\rm for}\quad 
\left|t_{{\rm obs,}z,i}-t_{z,j}+\frac{\hat{n}\cdot\vec{r}_{jk}}{c}\right|
<\frac{\Delta t_{{\rm obs,}z,i}}{2}\ .
\end{eqnarray}
\end{widetext}
Finally, one could simplify things even further by assuming an
isotropic emission in the fluid (comoving) rest frame, and then
$j'_{\nu'}(\hat{n}') = dE'/dV'd\Omega'd\nu'dt'$ can be replaced by
$P'_{\nu'}/4\pi$ where $P'_{\nu'} = dE'/dV'd\nu'dt'$. We currently
make this simplifying assumption.

For 2D jet simulations, which assume an axisymmetric flow, the jet
symmetry axis is the $z$-axis and it is convenient to choose the
$x$-axis along the $\hat{n}$-$\hat{z}$ plane, so that $\hat{n}$ may be
easily expressed in terms of the viewing angle $\theta_{\rm obs}$
(where $\cos\theta_{\rm obs} = \hat{n}\cdot\hat{z}$),
\begin{equation}
\hat{n} = \hat{x}\sin\theta_{\rm obs} + \hat{z}\cos\theta_{\rm obs}\ .
\end{equation}
Thus, in spherical $(r,\theta,\phi)$ or cylindrical $(z,\rho,\phi)$
coordinates (with $\beta_\phi=0$), we have
\begin{eqnarray}\nonumber
\hat{n}\cdot\vec{r} &=& 
r(\sin\theta\cos\phi\sin\theta_{\rm obs}+\cos\theta\cos\theta_{\rm obs})=\nonumber\\
 &=& \rho\cos\phi\sin\theta_{\rm obs}+z\cos\theta_{\rm obs}\ ,
\\ \label{eq:hats}
\hat{n}\cdot\vec{\beta} &=& 
(\beta_r\sin\theta+\beta_\theta\cos\theta)\cos\phi\sin\theta_{\rm obs} + \nonumber\\
&+&(\beta_r\cos\theta-\beta_\theta\sin\theta)\cos\theta_{\rm obs}
\\ \nonumber
&=&\beta_\rho\cos\phi\sin\theta_{\rm obs}+\beta_z\cos\theta_{\rm obs}\ .
\end{eqnarray}
%

\subsection{Calculation of the observed image}

The observed image can be calculated by dividing the plane of the sky
(i.e. the plane normal to $\hat{n}$) into bins or 2D ``pixels'' and
assigning the contribution $\Delta F_{\nu,jk}$ from each computational
4D cell to the appropriate pixels (or pixel), where the
conversion from flux to specific intensity (which is relevant for the
image calculation) is done by using the relation $dF_\nu = I_\nu
dS_\perp/d_A^2$, so that the intensity contribution to the $l^{\rm
th}$ pixel whose area is $\Delta S_{\perp,l}$ would be
\begin{eqnarray}
\Delta I_{\nu,il,jk}(\hat{n}) &=& d_A^2\frac{\Delta F_{\nu,i,jk}(\hat{n})}{\Delta S_{\perp,l}}=\nonumber \\
&=&\frac{(1+z)^{-2}\Delta V^{(4)}_{jk}}{\Delta S_{\perp,l}\Delta t_{{\rm obs,}i}}
\frac{j'_{\nu',jk}}{\Gamma^2_{jk}(1-\hat{n}\cdot\vec{\beta}_{jk})^2}= \nonumber \\
&=&\frac{(1+z)^{-3}\Delta V^{(4)}_{jk}}{\Delta S_{\perp,l}\Delta t_{{\rm obs,}z,i}}
\frac{j'_{\nu',jk}}{\Gamma^2_{jk}(1-\hat{n}\cdot\vec{\beta}_{jk})^2}\ .
\end{eqnarray}
The assignment of the contribution to the appropriate pixel requires a
parameterization of the plane of the sky. For this purpose we use a
rotated reference frame denoted by a twiddle, where $\tilde{y} = y$ and
the $\tilde{z}$-axis points to the observer (in the direction of
$\hat{n}$),
\begin{eqnarray}
\tilde{x} &=& x\cos\theta_{\rm obs}-z\sin\theta_{\rm obs}= \nonumber\\
&=& r(\sin\theta\cos\phi\cos\theta_{\rm obs}-\cos\theta\sin\theta_{\rm obs})=\nonumber\\
&=& \rho\cos\phi\cos\theta_{\rm obs}-z\sin\theta_{\rm obs}\ ,
\end{eqnarray}
\begin{eqnarray}
\label{eq:twiddle}
\tilde{y} = y = r\sin\theta\sin\phi = \rho\sin\phi\ , \qquad
\tilde{\rho} = \sqrt{\tilde{x}^2+\tilde{y}^2}\ ,
\end{eqnarray}
\begin{eqnarray}\nonumber
\tan\tilde{\phi} &=& \frac{\tilde{y}}{\tilde{x}} =
 \frac{\sin\theta\sin\phi}{\sin\theta\cos\phi\cos\theta_{\rm obs}-\cos\theta\sin\theta_{\rm obs}} =
\nonumber \\
&=&\frac{\rho\sin\phi}{\rho\cos\phi\cos\theta_{\rm obs}-z\sin\theta_{\rm obs}}\ .
\end{eqnarray}
For an axisymmetric flow the image is invariant to $\tilde{y} \to
-\tilde{y}$ or equivalently to $\tilde{\phi}\to-\tilde{\phi}$,
i.e. $I_\nu(t_{\rm obs},\hat{n},\tilde{x},\tilde{y}) = I_\nu(t_{\rm
obs},\hat{n},\tilde{x},-\tilde{y})$ and $I_\nu(t_{\rm
obs},\hat{n},\tilde{\rho},\tilde{\phi}) = I_\nu(t_{\rm
obs},\hat{n},\tilde{\rho},-\tilde{\phi})$.\footnote{This can also be
seen from equation~(\ref{eq:hats}), where the dependence on $\phi$ is only
through $\cos\phi$, which is invariant to $\phi\to-\phi$ that
according to equation~(\ref{eq:twiddle}) corresponds to $\tilde{y} \to
-\tilde{y}$ or $\tilde{\phi}\to-\tilde{\phi}$.} A  2D
simulation (whether in spherical or cylindrical coordinates) provides
2D snapshots of the dynamics, and each  2D computational cell (not
counting the time dimension) needs to be transformed into one or more
3D cells. For the special case of an observer along the jet (or flow)
symmetry axis, corresponding to $\theta_{\rm obs}=0$, the contribution
to the observed emission (i.e. to $I_\nu$ or $F_\nu$) becomes
independent of $\tilde{\phi}$, which in this case is equal to $\phi$,
so that the image has circular symmetry ($I_\nu$ becomes independent
of $\tilde{\phi}$) and a single bin in $\phi$ becomes sufficient for the calculation. For
$\theta_{\rm obs}>0$, however, one needs to artificially produce a
large number of bins in $\phi$, each corresponding to a 3D cell, which
together represent a single 2D computational region. The choice of
binning should be done wisely, such that the Doppler factor does not
vary by a large factor between neighboring bins (in order to calculate
the observed radiation accurately enough) and the bin size should not be too
coarse (as to cause excessive graininess in the calculated images or
lightcurves), while having a reasonable number of bins (in order for
the computational time not to be too large, especially for
high-resolution simulations). Please note that since the contribution
to the flux is invariant to $\phi\to-\phi$, it is enough to choose
values in the range $0<\phi<\pi$ and give each resulting 3D or 4D cell
a double weight when calculating $F_\nu$ (since $\phi_1<\phi<\phi_2$
also represents $-\phi_2<\phi<-\phi_1)$.

\subsection{Synchrotron radiation}

The main purpose of the current radiation calculations is to
check the effect of the dynamics on the afterglow lightcurves. Because of this, we intentionally
use a very simple model for the radiation mechanism \citep[following][]{GPS99a},
which features synchrotron emission and ignores inverse Compton
scattering or its effects on the synchrotron emission through the
additional electron cooling that it causes. It also ignores
self-absorption, and the local emission spectrum is approximated by a
broken power-law. The magnetic field is assumed to hold everywhere a
fraction $\epsilon_B$ of the proper internal energy density, $e'$,
i.e. $B'^2/8\pi = \epsilon_B e'$. Just behind the shock all electrons
are assumed to be accelerated into a power-law energy distribution,
\begin{equation}
N(\gamma_e)\propto \gamma_e^{-p}\quad {\rm for}\quad \gamma_e>\gamma_m
= \fracb{p-2}{p-1}\frac{\epsilon_e e'}{n'_em_ec^2}\ .
\end{equation}
The local emissivity $P'_{\nu'}$ is taken to be a broken power-law,
\begin{equation}
\frac{P'_{\nu'}}{P'_{\nu',{\rm max}}} = \left\{\matrix{
(\nu'/\nu'_m)^{1/3} & \nu'<\nu'_m<\nu'_c\ , \cr & \cr
(\nu'/\nu'_c)^{1/3} & \nu'<\nu'_c<\nu'_m\ , \cr & \cr
(\nu'/\nu'_m)^{(1-p)/2} & \nu'_m<\nu'<\nu'_c\ , \cr & \cr
(\nu'/\nu'_c)^{-1/2} & \nu'_c<\nu'<\nu'_m\ , \cr & \cr
(\nu'/\nu'_m)^{(1-p)/2}(\nu'/\nu'_c)^{-1/2} & \nu'>\max(\nu'_m,\nu'_c)\ ,
}\right.
\label{eq:nupl}
\end{equation}
with the following flux normalization and break frequencies,
\begin{eqnarray}
P'_{\nu',{\rm max}} &=& 
0.88\,\frac{512\sqrt{2\pi}}{27}\fracb{p-1}{3p-1}
\frac{q_e^3}{m_ec^2}(\epsilon_Be')^{1/2}n'_e\ ,
\\
\nu'_m &=& \frac{3\sqrt{2\pi}}{8}\fracb{p-2}{p-1}^2\frac{q_e}{m_e^2c^5}
\epsilon_B^{1/2}\epsilon_e^2(e')^{5/2}(n'_e)^{-2}\ ,
\\ \label{eq:nu_c}
\nu'_c &=& \frac{27\sqrt{2\pi}}{128}\frac{q_e m_e c}{\sigma_T^2}
(\epsilon_Be')^{-3/2}\fracb{\Gamma}{t_z}^2\ .
\end{eqnarray}
Electron cooling is treated in an approximate manner, by assuming that
everywhere the electrons have cooled at their current local cooling
rate over the dynamical time, which is in turn approximated as
$t'_{\rm dyn}\approx t_z/\Gamma$, so that the expression in
equation~(\ref{eq:nu_c}) is simply derived from
\begin{equation}
\gamma_c = \frac{3m_e c}{4\sigma_T\epsilon_Be't'_{\rm dyn}} \approx
\frac{3m_e c\Gamma}{4\sigma_T\epsilon_Be't_z}\ ,\quad\quad
\nu'_c = \frac{3q_e B'\gamma_c^2}{16m_e c}\ .
\end{equation}
A more proper treatment of the electron cooling would require
following each fluid element from the point where it crosses the shock
and the electrons are accelerated, and solving the equation for the
subsequent evolution of their energy distribution, accounting for
their radiative losses and adiabatic gains or losses. This has been
done analytically for the BMK self-similar solution \citep{GS02} and
numerically using a 1D Lagrangian code \citep{NG07}. It has also been
implemented in an Eulerian code \citep{vanEerten10a,vanEerten10b}, in
a somewhat approximate fashion due to the difficulty in accurately
tracking the electron energy distribution in each fluid element.  
The differences between our treatment of the electron cooling and
the results presented by \citep{GS02} are shown in detail in the 
Appendix.

It is also possible to use an even simpler emission model that ignores
electron cooling altogether by assuming $\nu',\nu'_m <\nu'_c$ in the
broken power-laws of equation (\ref{eq:nupl}).  In this paper electron
cooling is always implemented in our calculations.  In an accompanying
paper \citep{paperII}, however, in some cases we also use an even
simpler emission model that ignores electron cooling altogether,
\begin{equation}
\frac{P'_{\nu'}}{P'_{\nu',{\rm max}}} = \left\{\matrix{
(\nu'/\nu'_m)^{1/3} & \nu'<\nu'_m\ , \cr & \cr
(\nu'/\nu'_m)^{(1-p)/2} & \nu'>\nu'_m\ .
}\right.
\label{eq:nocool}
\end{equation}

\section{Application: evolution of a relativistic impulsive blast wave}
\label{sec4}

In this section, we use our AMR+radiation code to study the evolution
of impulsive relativistic blast waves both in one-dimension (1D) -- a
spherical blast wave propagating into either a uniform or a stratified
medium, bridging from the Blandford-McKee to the Sedov-Taylor
(ST) self-similar solutions -- and in two dimensions (2D) -- an
axi-symmetric jet propagating into a uniform medium.

\subsection{Self similar solution}

\citet[]{BM76} studied the self-similar 
propagation of an ultra-relativistic spherical impulsive blast 
wave in a medium with a density 
\begin{equation}
 \label{eq:bmdens}
  \rho_k(r) = A_k r^{-k} \ .
\end{equation}

They showed that an appropriate choice of the similarity variable is
\begin{equation}
\chi=1+2 (4-k) \Gamma_{\rm sh}^2\left(1-\frac{r}{R}\right)\; ,
\end{equation}
where $r$ and $R$ are the radial position (in polar coordinates) of
the fluid element and of the shock front respectively; $\Gamma_{\rm
sh}$ is the Lorentz factor of the shock front, which as that of the
fluid (and all of the velocities) here is measured in the rest frame of
the upstream medium ahead of the shock, and it is related to the the
Lorentz factor of the shocked fluid just behind the shock front by
$\Gamma(\chi=1) =\Gamma_{\rm sh}/\sqrt{2}$.
\citet[]{BM76} showed that the position of the shock front is given by
\begin{equation}
R = c t\left(1-\frac{1}{2(4-k) \Gamma_{\rm sh}^2} \right)\ ,
\label{eq:r}
\end{equation}
 and its Lorentz factor can be written as
\begin{equation}
\Gamma_{\rm sh}^2 = \frac{(17-4k)E}{8\pi \rho_k(R) c^5 t^3} \ ,
\label{eq:gamma}
\end{equation}
where $\rho_k(R) = A_k R^{-k}$ is the density of the ambient
(un-shocked) medium at the position of the shock front, and
$E$ is the energy in the blast wave.

The lab frame time corresponding to a given Lorentz factor of the
shock front is therefore (see equations (\ref{eq:bmdens}), (\ref{eq:r}) and
(\ref{eq:gamma})) given by
\begin{equation}
t \cong \frac{R}{c} \cong \frac{1}{c} \left[\frac{(17-4k)E}{8 \pi A_k c^2
\Gamma_{\rm sh}^2} \right]^{1/(3-k)} \ .
\label{eq:t}
\end{equation}

The post-shock Lorentz factor $\Gamma$, proper rest-mass density
$\rho$, and pressure $p$, are given by
\begin{eqnarray}
\label{eq:bm1}
\Gamma &=& \frac{1}{\sqrt{2}} \Gamma_{\rm sh} \chi^{-1/2}\ ,
\\
\label{eq:bm2}
\rho &=&  2^{3/2} \rho_k(R)\Gamma_{\rm sh}\chi^{-(10-3k)/[2(4-k)]}\ , 
\\ \label{eq:bm3}
p &=& \frac{2}{3}\rho_k(R)c^2\Gamma_{\rm sh}^2\chi^{-(17-4k)/[3(4-k)]}\ .
\end{eqnarray}

The relativistic blast wave typically begins to slow down when it sweeps up 
an amount of mass with a rest-mass energy of order of the
kinetic energy of the blast wave. That corresponds to a distance
(Sedov length) of
\begin{equation}
L_s = \left[\frac{(3-k)E}{4\pi A_k c^2}\right]^{1/(3-k)}\ ,
\label{eq:sedov}
\end{equation}
where the jet energy $E$ is the energy (excluding rest energy)
in the flow.  For a non-spherical flow, or a jet, to zeroth
order $E$ in equations (\ref{eq:gamma}) and (\ref{eq:t}) can be
replaced by the local value of the isotropic equivalent energy in the
flow, $E_{\rm iso} = 4\pi(dE/d\Omega)$, as long as it does not vary
significantly over an angular scale of the order of the inverse of the
local value of the Lorentz factor of the fluid just behind the
shock. In particular, for a double-sided conical wedge of half-opening
angle $\theta_0$ taken out of the BMK solution (or a uniform
sharp-edged jet), which we later use as the initial conditions of our
2D simulations, we have $E = (1-\cos\theta_0)E_{\rm iso}\approx
(\theta_0^2/2)E_{\rm iso} \approx 2\times
10^{51}(\theta_0/0.2)^2E_{\rm iso,53}\;$erg, where we have used a
fiducial value of $E_{\rm iso} = 10^{53}E_{\rm iso,53}\;$erg, typical
for long duration GRBs. Whether it is more appropriate to use $E$ or
$E_{\rm iso}$ in equation (\ref{eq:sedov}) for such a jet, i.e. at
which distance from the origin it becomes Newtonian, is a non-trivial
question, which is addressed in an accompanying paper \citep{paperII}.

In the non-relativistic limit, the self-similar behavior of the blast
wave is described by the Sedov-Taylor \citep{Sedov59, Taylor50}
self-similar solution, with the position of the shock wave given by
\begin{equation}
 R \approx \left[\frac{\alpha_k E_{\rm iso} t^2}{A_k}\right]^{1/(5-k)}\ ,
\label{eq:nr}
\end{equation}
and the shock velocity given by $v_{sh} = dR/dt \propto t^{-(3-k)/(5-k)}$. 
Approximated expressions for the post-shock density, pressure and velocity
profiles in the ST regime are given, e.g., by \citet{2000A&A...357..686P}.
As there are not analytical solutions for the scaling
of density, pressure and velocity in the post-shock region, it is not possible to find
a simple analytical expression for $\alpha_k$. Based on the simulations presented in \S \ref{sec:ic}), we find $\alpha_k^{1/(5-k)}=1.15, 1.04, 0.78$ 
for $k=0,1,2$ respectively.

\subsection{Initial conditions}
\label{sec:ic}

In this paper, we perform a series of 1D (with $k=0,1,2$) and 2D (with $k=0$)
simulations of the propagation
of impulsive blast waves, including the transition from the relativistic
to the non-relativistic phase. All simulations employ spherical 
(polar) coordinates,
and using the HLL method (see \S \ref{sec:int}) for the flux calculation.
The multi-dimensional simulations for the cases $k=1,2$ are 
presented in an upcoming paper \citep{paperII}.

The initial conditions of the problem depend on the values of the
following parameters: the isotropic energy of the blast wave, $E_{\rm
iso}$, the initial Lorentz factor of the jet shock front, $\Gamma_{\rm
sh,0}$, the density profile of the external medium (that is, the values
of $k$ and of the normalization factor $A_k$) and the jet initial
half-opening angle, $\theta_0$ (in the 2D case).
In all the simulations, the initial profiles of density, pressure and
Lorentz factor (radial velocity) in the post-shock 
region are set from the BMK self-similar solutions, given by 
equations~(\ref{eq:bm1})-(\ref{eq:bm3}). 
We initialize the density of the ambient medium 
(in the case $k=0$) as $A_0 = \rho_0 = n_0 m_p = 1.67\times 10^{-24}\;{\rm g\;cm^{-3}}$, 
and the pressure as $p = \eta \rho_0 c^2$, 
with $\eta=10^{-10}$. The value of $\eta$ does not affect 
the outcome of the simulation as long  as the Mach number remains 
large, i.e. $\mathcal{M} \sim \eta^{-1/2} v_{\rm sh}/c \gg 1$.  
As the simulation continues to evolve well into the Newtonian regime, this
condition corresponds to $v_{\rm sh}
\gg 3\; (\eta/10^{-10})^{1/2}\;{\rm km\;s^{-1}}$.

In a first set of simulations, we study the deceleration 
of mildly relativistic impulsive blast waves bridging from the BMK
to the ST self-similar solution. In the case $k=0$, the initial
conditions are similar to those used by \citet{vanEerten10a}.
To determine the density profile in the cases $k=1,2$, we fix the 
Sedov length (equation \ref{eq:sedov}) as $L_s(k)=L_s(k=0)$:
\begin{equation}
  L_{\rm s} = \left[\frac{(3-k)E}{4\pi A_k c^2}\right]^{1/(3-k)} = 
        \left[\frac{3 E}{4\pi A_0 c^2}\right]^{1/3} \ ,
  \label{eq:deck}
\end{equation}
and derive an expression for $A_k$ as
 $ A_k = A_0 L^k_{\rm s} (3 - k)/3 $.
Therefore
\begin{equation}
  \rho = A_0 \frac{3-k}{3} \left(\frac{r}{L_{\rm s}}\right)^{-k} \; .
  \label{eq:dec0}
\end{equation}

We further assume $E_{\rm iso} = 10^{52}$~ergs, corresponding to a
Sedov length of $L_s= 1.17 \times 10^{18}\;$cm, and a Lorentz factor
of the shock of $\Gamma_{\rm sh,0} = 10$.  To properly cover the
deceleration to non-relativistic speeds (especially for the case
$k=2$), we use a large spherical box of radial size $L_r = 3\times
10^{20}\;$cm (corresponding to a size of $\approx 256 L_{\rm s}$).  The simulation is stopped
at $t_{\rm fin}= 500$~yrs.

In the case $k=0$, the simulations begins at 
$t_0 = 1.19 \times 10^7$~s, with a 
jet shock located at $R_0 = 3.56 \times 10^{17}$~cm. 
The case $k=1$ corresponds to an initial time and jet shock radius
given by $t_1/t_0 = R_1/R_0 = 0.53$.  The case $k=2$, corresponding to
a steady spherically symmetric wind, has $t_2/t_0 = R_2/R_0 = 0.074$.
The values assumed for the spherical wind can be compared with those
observed for Wolf-Rayet stars, which winds have large mass-loss rates
of $\dot{M} \approx 10^{-5}-10^{-4}$~M$_\sun$yr$^{-1}$ and velocities
$v_{\rm w} \approx 1000-2500$~km~s$^{-1}$ \citep[e.g.,][]{1986ARA&A..24..329C},
giving
 $n_{\rm w}(r) \approx 0.45
    (r/10^{18}\,\rm{cm}) ^{-2}
    (\dot{M}_{\rm wr}/3\times 10^{-5}\,M_\sun\,yr^{-1}) 
    (v_{wr}/2\times 10^3\,km\,s^{-1})\;\rm{cm}^{-3}$,
which is very similar to the one used in the simulations.

The AMR code uses a basic grid of 1000 cells with a maximum of 18 
levels of refinement, corresponding 
to a maximum resolution of $\Delta r_{\rm min} = 2.3 \times 10^{12}\;$cm. 
In a uniform grid code, the same resolution would be achieved by using 
$1.3\times 10^8$ cells.

In a second set of simulations, we test the radiation code by 
running simulations of highly relativistic decelerating blast waves
(limited to the case $k=0$) both in 1D and 2D.
In these simulations, we assume an isotropic energy of  
$E_{\rm iso} = 10^{53}$~ergs, corresponding to a Sedov 
length of $L_s = 2.51\times 10^{18}\;$cm, and a Lorentz factor 
of $\Gamma_{\rm sh,0} = \sqrt{2} \times 20$. 
The simulations begins at $t_0 = 1.277 \times 10^7$~s, 
with the shock initially located at $R_0 = 3.83 \times 10^{17}$~cm,
and ends at $t_{\rm fin}= 150$~yrs.
To properly study its lateral expansion, an initial opening 
angle of $\theta_0 = 0.2$~rad (in the 2D case) is assumed for the jet.

\begin{figure}
\centering
 \includegraphics[width=0.35\textwidth]{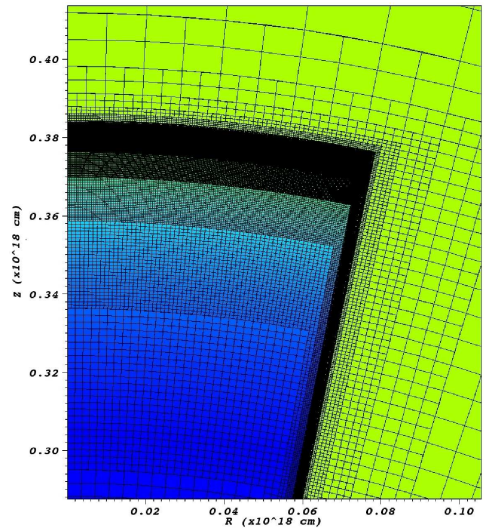}
\caption{Adaptive grid structure for the initial condition of the 
two-dimensional simulation of a relativistic blast wave. The blue
color indicates the post-shock region, while the green area
represent the ambient (unshocked) medium.}
\label{fig2}
\vskip 0.5cm
\end{figure}

The spherical box has a radial size of $L_r = 1.1\times 10^{19}\;$cm and
angular size (in the 2D simulation) $L_\theta = \pi/2$.
The AMR code uses a basic grid of 100 cells along the radial direction
both in 1D and 2D, and 4 cells
along the $\theta$ direction in the 2D
simulations. We run a series of simulations varying the maximum number
of refinement levels. The lowest resolution simulation uses 10 maximum levels, 
while the highest employs 18 levels of refinement in 1D and 
15 in 2D, corresponding to a maximum
resolution of $\Delta r_{\rm min} = 2.1 \times 10^{11}\;$cm in 1D and
$\Delta r_{\rm min} = 6.7 \times 10^{12}\;$cm, $\Delta\theta_{\rm min}
= 2.4\times10^{-5}\;$rad (along $r$ and $\theta$) in 2D.  The
structure of the grid at the beginning of the simulation is shown in
Figure~\ref{fig2} for the 2D run.  In a uniform grid code, the same
resolution would be achieved by employing $5.2\times 10^7$ cells in 1D,
and $\sim 10^{11}$ cells in 2D.


To keep approximately constant the resolution of the 
relativistic thin shell $\Delta \propto t^{4-k}$, the maximum 
number of levels of refinement $N_{\rm levels}$ is decreasing with 
time \citep{Granot07} as $N_{\rm levels} = \max[7, N_{\rm levels, 0} -
(4-k) \log(t/t_0)/\log(2)]$. 
We refine our adaptive mesh based on rest mass density and energy gradients.
The 1D simulations run in at most a few hours on a normal workstation,
while the 2D simulations need a few days on $\sim 100$ processors.

\subsection{One-dimensional simulations of trans-relativistic 
blast waves propagating in a stratified medium ($k=0,1,2$)}


\begin{figure}
\centering
 \includegraphics[width=0.5\textwidth]{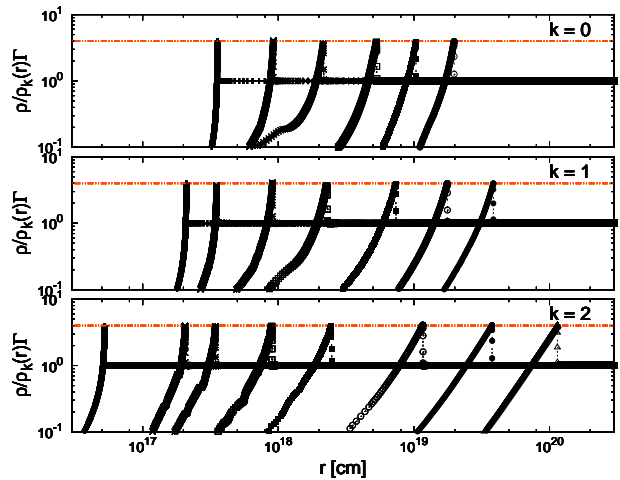}
\caption{Density profiles normalized to the Lorentz factor and the 
local value of the ambient medium density. 
The curves shown in the upper panel ($k=0$) correspond to 
$t=t_0=137$~days, and $t=1, 3, 20, 100, 500$~yrs. The central panel
($k=1$) includes also the profile at $t_0=81$~days. The bottom panel
($k=2$) includes also the profile at $t_0=20.3$~days. The horizontal 
red line indicates $\rho/\rho_k(r)\Gamma=4$.}
\label{fig3}
\vskip 0.5cm
\end{figure}

During its deceleration, the shock front is typically resolved with
3-4 cells (Figure~\ref{fig3}), as is the case for most modern Eulerian shock
capturing schemes.  The normalized lab frame density behind the shock,
given from the relativistic Rankine-Hugoniot conditions for strong
shocks: $\rho/\rho_k(R) \Gamma =
(\bar\gamma+1/\Gamma)/(\bar\gamma-1)$, remains approximately constant
during the transition from relativistic to non-relativistic regimes
\citep[see, e.g.,][]{BU06,vanEerten10a}. Figure~\ref{fig3} shows that in
fact $\rho/\rho_k(R) \Gamma \approx 4$ at different times and for
different values of $k$.  The drop of the density profile in the
post-shock region approximately follows the BMK self-similar solution, and is
therefore less steep with larger $k$ (see equation \ref{eq:bm2}).
This figure also shows that the deceleration process is slower for a
more stratified medium.


\begin{figure}
\centering
 \includegraphics[width=0.45\textwidth]{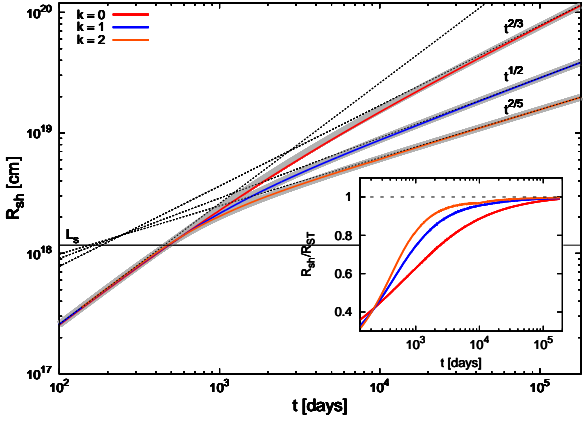}
\caption{
Position of the shock front for the three cases
$k=0, 1, 2$ (up to bottom) along with the 
ultra-relativistic ($R_{sh} = c t$)
and the Sedov-Taylor ($ R_{\rm sh} = 
\left(\alpha_k E_{\rm iso} t^2/A_k\right)^{1/(5-k)}$) regimes.
The Sedov-Taylor curves assume $\alpha_k^{1/(5-k)}=1.15, 1.04, 0.78$ for
$k=0,1,2$ respectively. 
The gray curves are computed from a semi-analytical
approximation based on energy conservation (see the text for
a detailed description).
}
\label{fig4}
\vskip 0.5cm
\end{figure}

Figure~\ref{fig4} shows the evolution of the 
shock front radius for different density stratifications.
Both the ultra-relativistic (with $R_{\rm sh} \approx c t$)
and the non-relativistic ($R\propto t^{2/(5-k)}$) 
analytical self-similar solutions are 
properly recovered by the simulations.
As shown e.g. by \citet{vanEerten10a} for the case $k=0$, 
the transition from relativistic to non-relativistic 
phase happens on scales much larger than $L_s$.
If for instance we estimate from Figure~\ref{fig4} the time it takes 
for the relativistic blast wave to slow down to 
non-relativistic speeds based on the intersection between
the relativistic and non-relativistic self-similar
curves, we obtain values of $\sim 0.9\times 10^3$~days, 
$1.2\times 10^3$~days and $2.7\times 10^3$~days for 
$k=0,1,2$ respectively. These values are much larger
than those computed by using the Sedov length (Piran 2005)
$t_{\rm NR}\sim L_{\rm NR}/c = 450$~days.

This result, together with the scaling of position, 
Lorentz factor and the shock velocity as a function of time and
shock radius, can be easily understood
by a simple analytical argument involving the conservation of energy.
In fact, the energy is given in the 
ultra-relativistic regime by
\begin{equation}
   E = \frac{8 \pi}{17 - 4 k} A_k c^2 R^{3-k} \Gamma^2 \beta^2
\end{equation}
and in the non-relativistic limit by
\begin{equation}
   E = \frac{(5-k)^2}{4 \alpha_k} A_k v^2 R^{3-k}
\end{equation}

As the energy has a common scaling in relation with the  other physical 
parameters ($v$,$R$), differing only in the constant of proportionality,
a simple interpolation between the two limits is given by
\begin{equation}
  E = R^{3-k} \beta^2 \Gamma^2 A_k c^2 \left( \frac{8 \pi}{17 - 4 k} \beta^2 
          + \frac{(5-k)^2}{4 \alpha_k} (1- \beta^2)\right)
\end{equation}

This equation can be easily written as function of velocity
as:
\begin{equation}
\beta^2 = \frac{2}{1 + c_{NR}(R/L_s)^{3-k} + \sqrt{[1-c_{\rm NR}(R/L_{\rm s})^{3-k}]^2 + 4 c_R (R/L_{\rm s})^{3-k}}} \ ,
\label{eq:fit}
\end{equation}
where $c_R = \frac{2(3-k)}{17 - 4 k}$ and $c_{\rm NR} =
\frac{(5-k)^2(3-k)}{16 \pi \alpha_k}$. This expression approximately
gives the dependence of $v$ (or $\Gamma$) on the shock position, for
every choice of the blast wave energy and density stratification. For
instance, at $R \sim L_{\rm s}$, equation (\ref{eq:fit}) gives $v_{\rm
sh} \sim 0.83,0.85,0.89 c$ (or $u = \Gamma\beta\sim 1.46,1.64,1.99$)
for $k=0,1,2$ respectively.  At this radius (and time) the shock is
therefore still relativistic, and the ST solution is not
valid. 
The exact determination of $t_{\rm NR}$ depends, however, on the definition of the 
transition between the relativistic and the non-relativistic flow \citep[e.g.][]{2010ApJ...716.1028R}. If
for instance we define $t_{\rm NR}$ as the time where 
the asymptotic BMK solution and the ST power laws are equal (i.e. $ct/L_s =
[4\pi\alpha_k/(3-k)]^{1/(3-k)}$), we get $t\sim 2.1
t_{\rm NR} \sim 9\times 10^2$~days ($k=0$), $t\sim 3 t_{\rm NR} \sim
3.4$~yrs ($k=1$) and $t\sim 6 t_{\rm NR}\sim 7.5$~yrs ($k=2$).
At this time the blast wave is nonetheless  still mildly relativistic 
($\beta=0.51, 0.56, 0.63$) and the ST solution is not valid.
If on the other hand we assume that the ST solution becomes valid at
a fixed (somehow arbitrary) speed of $v/c\lesssim 1/3$, we get $t\sim 3.6
t_{\rm NR} \sim 1.6\times 10^3$~days ($k=0$), $t\sim 7 t_{\rm NR} \sim
8.6$~yrs ($k=1$) and $t\sim 48 t_{\rm NR}\sim 59$~yrs ($k=2$) (Figure~\ref{fig4}).

Equation (\ref{eq:fit}), when rewritten 
in the form $dR/dt = \beta(R)$, admits a complex solution $t=f(R)$
in terms of Appell hypergeometric functions.
The time dependence of the shock position $R = R(t)$ has been therefore 
more easily derived by numerical integrating equation (\ref{eq:fit}), 
and it approximates the position of the shock computed from the numerical 
simulation within a maximum difference of 1, 2, 5\% (for $k=0,1,2$ respectively).

While Figures~\ref{fig3} and \ref{fig4} clearly show the validity
of our implementation for mildly-relativistic and non-relativistic
speeds, reproducing the correct BMK
self-similar scaling during the early stages of the simulation,
when $\Gamma \gtrsim 10$, is much more challenging.

\begin{figure}
\centering
 \includegraphics[width=0.5\textwidth]{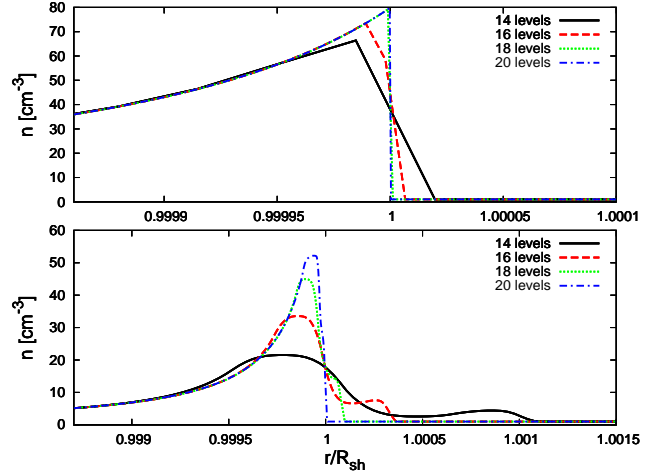}
\caption{
Number density profile (in the lab frame) for different
resolutions for the case $k=0$ at the beginning of the
simulation ($t=148$~days ,upper panel)
and $t=156$~days (bottom panel).
}
\label{fig5}
\vskip 0.5cm
\end{figure}

Figure~\ref{fig5} shows the initial density profile (for the case
$k=0$) in the region around the position of the shock.
A very large number of levels of refinement must be used to 
properly initialize the density, pressure and Lorentz factor
in the post-shock region. For instance (Figure~\ref{fig5}, 
upper panel), the initial steep density profile is recovered 
with errors less than 10\% only by using resolutions corresponding
to $\gtrsim 18$ levels of refinement.


\begin{figure}
\centering
 \includegraphics[width=0.45\textwidth]{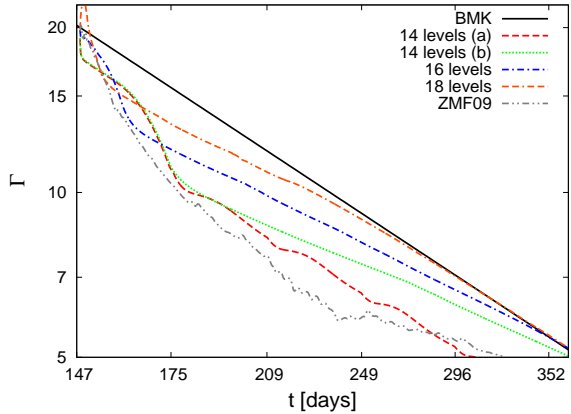}
\caption{
Maximum Lorentz factor in the post-shock region (measured in the lab
frame) as a function of time.  The simulations (with $k=0$) start at
$t\sim 147$~days with a Lorentz factor of 20. The curves shown
correspond to the expected BMK self-similar solution, 14,
16, 18 levels of refinement with a fixed maximum level of refinement,
and results of the \citet{ZM09} two-dimensional
simulations. For 14 levels of refinement we show two curves, either
with (a) or without (b) decreasing the maximum level of refinement
with time.  Each label on the $x-$axis corresponds to the time when
the maximum resolution drops by a factor of two, so that for instance
the simulation with initially 14 levels of refinement drops to 13
levels after 175~days, 12 after 209~days, and so on.}
\label{fig6}
\vskip 0.5cm
\end{figure}

While the BMK self-similar solution represents an exact 
solution of the SRHD equations in the ultra-relativistic
limit, the particular discretization employed may not be
the exact (numerical) solution of the discretized equations.
As a consequence, the relaxation towards the numerical
solution passes through the development (see Figure~\ref{fig5}, bottom
panel) of a spurious numerical ``precursor'' propagating in front of
the BMK shock if  insufficient resolution is used.  
While the size of the precursor shock
drops effectively with resolution, it also produces a quick drop in the
maximum Lorentz factor behind the shock (due to the spreading of the
initial $\Gamma$ peak, see Figure~\ref{fig6}). The Lorentz factor
eventually converges to the correct BMK solution at $\Gamma \sim 10$
at the largest resolution used (18 levels of refinement).
Figure~\ref{fig6} also shows the effect of decreasing the maximum
level of resolution during the evolution of the simulation \citep{Granot01,
ZM09, paperII}. As can be
appreciated from Figure~\ref{fig6}, the decrease in the resolution
produces a slower convergence to the BMK solution. The
time evolution of $\Gamma$ from \citet{ZM09}, included in
Figure~\ref{fig6} is similar to our low resolution (14 levels) 
one-dimensional simulation, corresponding approximately to the
resolution achievable in multi-dimensional simulations.  
The noise in the \citet{ZM09} curve is due to a 
larger temporal sampling.
A proper treatment of the tiny ultra-relativistic post-shock region
would require a larger resolution or alternatively a much less diffusive 
method as e.g. high order (coupled to high resolution) or 
Lagrangian-Eulerian methods \citep[e.g.][]{KPS99}.


The specific  numerical resolution required is determined by  the relevant 
structure one needs to resolve. The hardest to resolve, in our case, is
the initial BMK shell ($\Delta$) at the initial time ($t_0$) or radius ($R_0$).
Its effective width does not have a unique definition, but it can be
parameterized as
\begin{equation}\label{eq:def-a}
\Delta_0 = a\,\frac{R_0}{\Gamma_{\rm sh}^2(R_0)}\ ,
\end{equation}
where the numerical factor $a$ can be evaluated using the BMK self
similar solution
\footnote{Note that if one uses $\Gamma(R_0,\chi=1)$
instead of $\Gamma_{\rm sh}(R_0)$ in equation~(\ref{eq:def-a}) then the
value of the numerical coefficient $a$ would be smaller by a factor of
2.}.
Defining $\Delta_0$ as the width of the region behind the shock
that contains a fraction $f$ of the total energy ($E$) or rest mass
($M$), respectively, results in
\begin{equation}
a = \frac{(1-f)^{-\alpha}-1}{2(4-k)}\ ,\quad
\alpha_E = \frac{3(4-k)}{17-4k}\ ,\quad
\alpha_M = \frac{4-k}{3-k}\ .
\end{equation}
For $f = 1/2$, this gives $a_E = 0.0789, 0.103, 0.147$ and $a_M =
0.190, 0.305, 0.750$ for $k = 0, 1, 2$. 

One can then similarly express the numerical resolution in terms of a
parameter $a_{\rm res}$,
\begin{equation}\label{eq:def-a_res}
\Delta r_{\rm min} = a_{\rm res}\,\frac{R_0}{\Gamma_{\rm sh}^2(R_0)}\ ,
\end{equation}
where $\Delta r_{\rm min}$ is the smallest resolution element in the
radial direction.  Previous 2D jet numerical simulations with similar
initial conditions used $k= 0$. In ~\citet{Granot01} the initial
resolution was rather poor, $a_{\rm res} = 0.69$, while in ~\citet{ZM09} it was
significantly improved, $a_{\rm res} = 0.12$. Here we use $a_{\rm res} =
0.014$ for $k=0$, which represents an order of magnitude
improvement. For $k = 1$ and $2$ we have $a_{\rm res}
= 0.022$ and $0.087$, respectively.

\begin{figure}
\centering
 \includegraphics[width=0.3\textwidth,angle=-90]{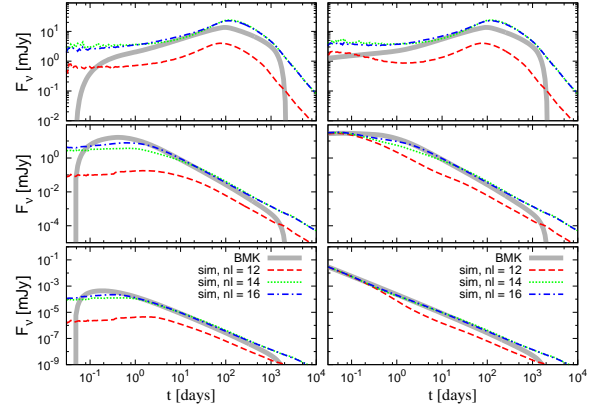}
\caption{
Light curves (at $10^{9/13/17}$~Hz from up to bottom) 
from simulations at different resolution, either
including (right panels) or not including (left panels)
the contribution from the synthetic lightcurve
emitted from a Blandford-McKee self-similar 
blast wave with Lorentz factor between 200 and 20.
The synthetic light curve (labeled BMK in the Figure)
emitted from a Lorentz factor between 1 and 20 (or 
between 1 and 200) is also shown in the left (right) 
panels of the figure.
}
\label{fig7}
\vskip 0.5cm
\end{figure}

Figure~\ref{fig7} shows that the light curve, computed by post-processing
the results of the simulations with our radiation code, converges
quickly except for $t_{\rm obs} \lesssim 0.5$~days,
where part of the flux, which should be generated
from regions with $\Gamma \sim 20$, is shifted to a lower $t_{\rm
obs}$. That can be in part compensated by adding the contribution
coming from the jet decelerating with $20 \leq \Gamma \leq 200$,
computed by mapping in the radiation code a
BMK self-similar solution.  As shown in Figure~\ref{fig7}
(right), the sum of the synthetic flux with $20 \leq \Gamma \leq
200$ and the flux computed from the results of the simulation
with $1 \leq \Gamma \leq 20$ produces a
valley (shallower for increasing resolutions) for $t_{\rm obs} \sim
1$~day. This artificial feature is due to relaxation from the
initial conditions to the numerical solution, and gradually disappears
as the resolution is increased. A comparison between the lightcurve 
computed from the 1D simulation (with $k=0$) and the 
semi-analytical calculations from \citet{GS02} is shown in the Appendix.


\begin{figure*}
\centering
 \includegraphics[width=0.8\textwidth]{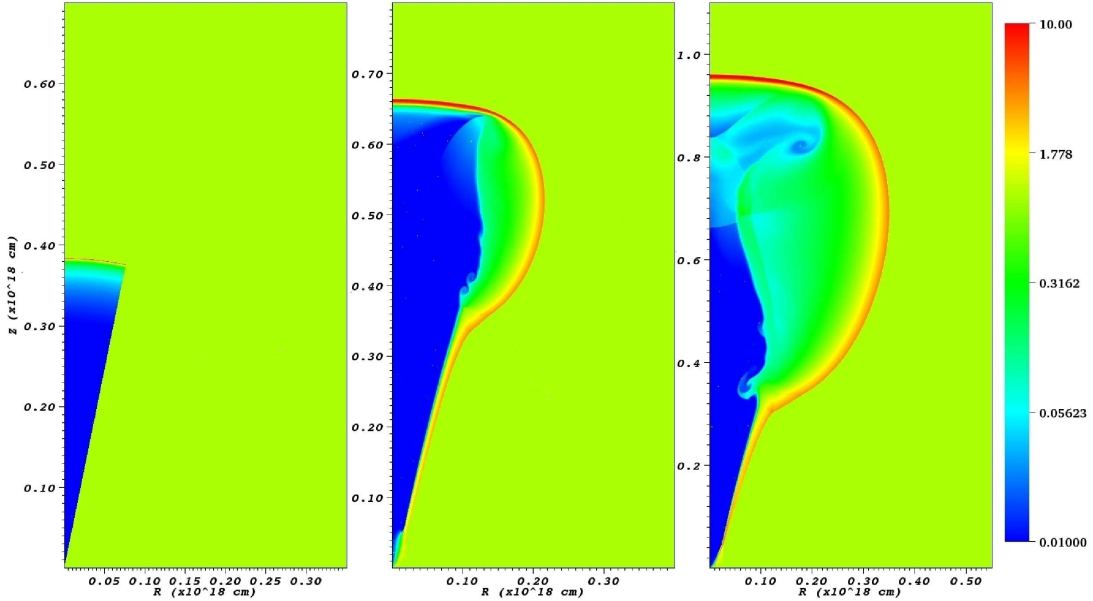}
\caption{
Lab-frame density stratification snapshots of the 2D simulation at 147~days
(left), 256~days (center), 372~days (right panel).}
\label{fig8}
\vskip 0.5cm
\end{figure*}

\subsection{Two dimensional simulations for $k=0$}

Figure~\ref{fig8} shows snapshots representing the early evolutionary
stages of the jet density.  During the relativistic phase, there is
only modest lateral expansion. As portions of the jet
expand laterally, a rarefaction front moves towards the jet axis.  The
strong shear present at the contact discontinuity drives
shearing instabilities that have however a negligible effect on
the shock dynamics and afterglow radiation coming from the jet. At
the jet break time $t=t_{\rm JB}\sim 8.7$~yr, the lateral expansion becomes more 
vigorous, and at later stages (on times $\gg t_{\rm NR}$) the jet slowly 
converges to a spherical shape.  
Although it is not possible to make a quantitative comparison,
our results qualitatively resemble those of \citet[][see their Figure 2 
for a direct comparison]{ZM09}, as well as those of
~\citet{Granot01}. 

While theoretical arguments \citep{2000astro.ph.12364G, 2002ApJ...568..830W} 
seems to indicate that the shock front should be stable to linear 
perturbations for either a uniform or a wind density profile of the
ambient medium, recent simulations by \citet{2010A&A...520L...3M} observe the 
development of instabilities in the shock front. The development of 
similar instabilities is also observed by \citet{paperII} relative 
to the case of a stratified medium with $k=2$, while it is not observed 
in the simulations presented in this paper (despite using the same HLL 
Riemann solver as \citealt{2010A&A...520L...3M} and similar initial conditions), 
consistently with the results by \citet{ZM09}.
The different results in the simulation seems to imply a numerical origin 
for the instabilities observed by \citet{2010A&A...520L...3M}, although 
further investigation is needed to better understand the problem.

The afterglow light curves computed from our 2D jet
simulation assume that the observer is located along the jet symmetry axis ($\theta_{\rm
obs}=0$). To facilitate comparison with the results of \citet{ZM09}, 
we choose the same parameters for the afterglow
calculation: $\epsilon_B = \epsilon_e = 0.1$, $z=1$ and $p=2.5$, in
addition to the same values for the parameters to determine the
hydrodynamics ($E_{\rm iso}=10^{53}\;$erg, $n_{\rm ext}=1\;{\rm
cm^{-3}}$ and $\theta_0=0.2\;$rad).


\begin{figure}
\centering
 \includegraphics[width=0.4\textwidth]{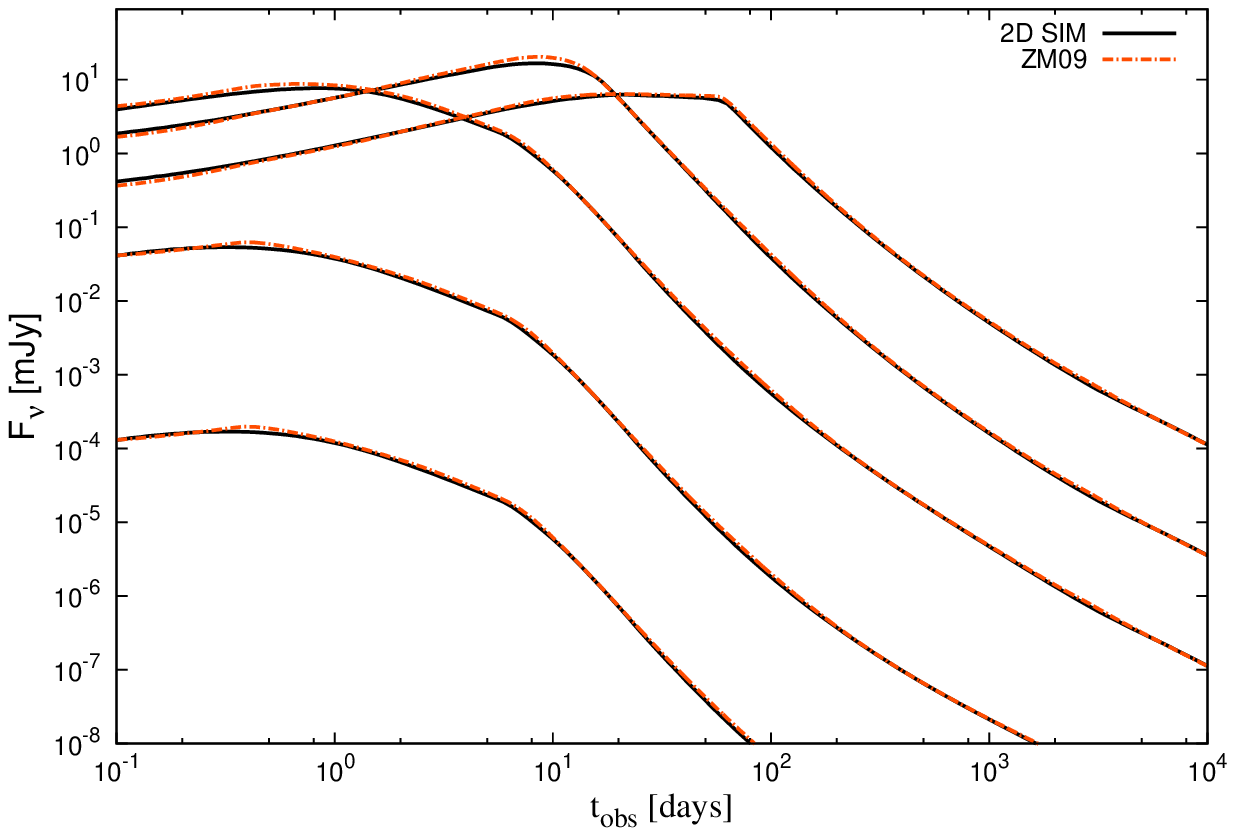}\\
 \includegraphics[width=0.4\textwidth]{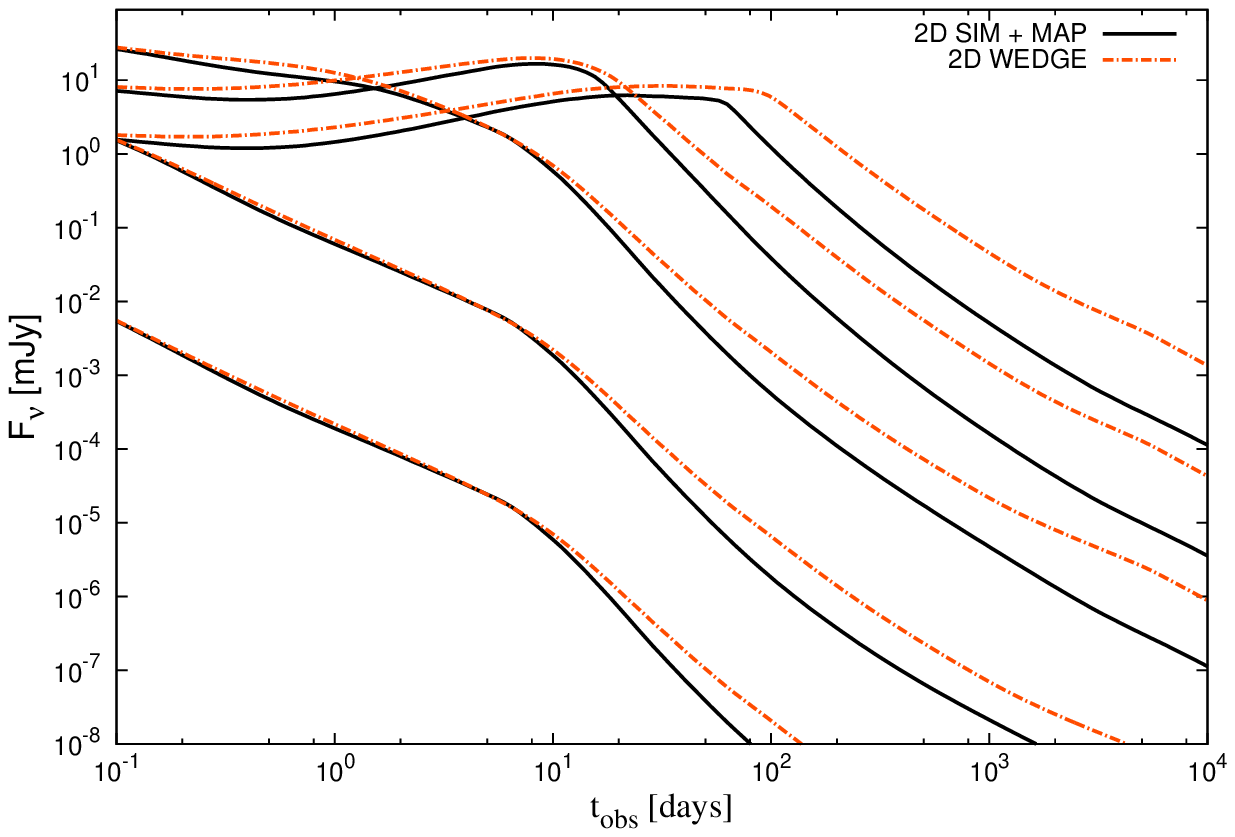}
\caption{Afterglow emission at $10^{9/11/13/15}$~Hz
from the 2D simulation compared with a 2D wedge (bottom panel)
and the results from \citet{ZM09} (upper
panel).
\label{fig9}}
\vskip 0.5cm
\end{figure}

\begin{figure}
\centering
 \includegraphics[width=0.45\textwidth]{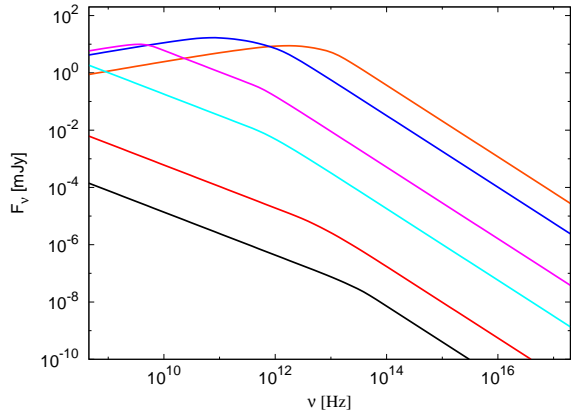}
\caption{Spectra at $t_{obs}=0.1,1,10,100,1000$~days
(black, red, green, blue, purple)
\label{fig10}}
\vskip 0.5cm
\end{figure}

As in the 1D case, the afterglow emission (Figure~\ref{fig9}) shows a
shallow valley at $t\lesssim 1$~day, due to a lack of resolution into
the region immediately behind the high relativistic shock. Figure~\ref{fig9} 
(bottom panel) shows a comparison with a 2D ``wedge''
(computed by using a 1D simulation mapped on a wedge with $\theta \leq
0.2$; the finite resolution of this 1D simulation is affecting
the lightcurves at the earliest times as shown in
Figure~\ref{fig9}). Before the jet break time, the 2D light curve from
the simulation is very similar to that from a 2D wedge with the same
(initial) isotropic energy, indicating that little sideways expansion
takes place before the jet break, in agreement with previous analytical
\citep[e.g.,][]{Rhoads99} and numerical \citep{Granot01} results. After
the jet break time, however, the flux from the 2D simulation becomes
lower than that for the corresponding wedge, and the difference
between the two gradually increases with time, as the lateral
spreading of the jet gradually increases during the relativistic phase
and then more rapidly during the Newtonian phase (until at very late
times spherical symmetry is approached).
Our calculated afterglow emission and spectra agree very well with \citet{ZM09}
(Figure~\ref{fig9}, upper panel and Figure ~\ref{fig10}) both in the flux 
before and after the jet break.


\section{Conclusions}
\label{sec5}

In this paper, we have presented a detailed description of the new
state-of-the-art adaptive mesh refinement, relativistic hydrodynamics 
code Mezcal-SRHD and of the radiation code used to compute the synchrotron 
emission from the output of the hydrodynamics simulation.  The proper
implementation of the SRHD algorithm has been verified by running
standard one- and multi-dimensional tests which are presented in the Appendix.  
The code has been applied to the study of the propagation of ultra-relativistic 
impulsive blast waves both in one and two-dimensional spherical coordinates. 

We have studied for the first time the deceleration of relativistic impulsive 
blast-waves in one dimension propagating in a stratified medium and find
that the deceleration to non-relativistic speeds happen on
scales $R_{\rm NR}$ from a few (for $k=0$) to several times larger 
than the Sedov length $L_{\rm s}$. Taking $R_{\rm NR}$ as the radius
where $R_{\rm ST}(t) = ct$ gives the expression $R_{\rm NR}/L_s =
[4\pi\alpha_k/(3-k)]^{1/(3-k)}$, which illustrates how 
$R_{\rm NR}/L_{\rm s}$ increases with the
degree of stratification of the ambient medium where the shock is
propagating. These results have been described in detail
using  a simple semi-analytical formula, derived from energy conservation,
which gives the correct scaling of the position and velocity of the shock
as a function of time.

The results obtained by the radiation code were validated by a
comparison with semi-analytical results, and with those obtained in
previous numerical works. We have also shown that while the resolution
is a key factor to properly recover the correct dynamical evolution of
the system (with some of the parameters not yet converging, e.g. the
shock Lorentz factor), when the contribution from the radiation
produced by the jet before the onset of the simulation (in our case
$20 \leq\Gamma_{\rm sh}/\sqrt{2} \leq 200$) is included in the calculation, 
the resulting light curve becomes much less sensitive to the exact resolution.

In an upcoming  paper, we will extend the results of the simulations 
presented here to include multi-dimensional simulations in a 
stratified medium. The study of the contribution of the magnetic 
field on the jet dynamics and afterglow radiation is left for future 
works.

\acknowledgements
This research was supported by the David and Lucille
Packard Foundation (ERR and FDC), the NSF (ERR) (AST-
0847563), the ERC advanced research grant ``GRBs''
and a DGAPA postdoctoral grant from UNAM (DLC).
We thank Weiqun Zhang to share with us data from 
his 2D simulations (used in Figure \ref{fig6}).

\appendix 
\label{appendix}

\section{Evaluating the approximations used in the electron cooling
frequency estimation}

\begin{figure}
\centering
 \includegraphics[width=\textwidth]{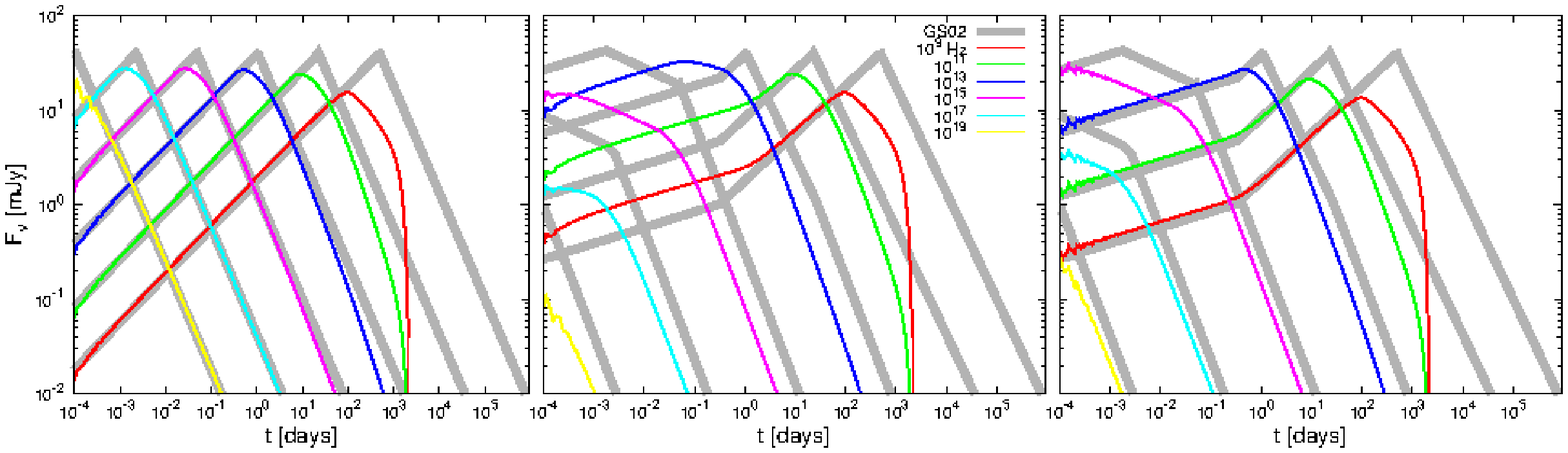}
\caption{
Comparison between  light curves (at
$10^{9/11/13/15/17}$~Hz) computed from a
Blandford-McKee self-similar blast wave with Lorentz factor between
600 and 1, and the semi-analytical results from \citep{GS02}.
\emph{Left panel}: Simple emission model excluding electron cooling 
(equation \ref{eq:nocool}).
\emph{Center}: Light curve computed by using an approximated 
emission model for the electron cooling (equations \ref{eq:nupl}).
\emph{Right}: The same as the center panel, but with a cooling 
frequency four times larger.
}
\label{fig11}
\vskip 0.5cm
\end{figure}

A comparison between the lightcurve computed by mapping 
in the radiation code a blast wave described by a BMK 
self-similar solution and the semi-analytical calculations from \citet{GS02}
is shown in Figure~\ref{fig11}. While \citet{GS02} obtained
smooth spectral breaks, for simplicity we use here their broken
power-law prescription (without synchrotron self-absorption). In that work the
afterglow emission from the BMK solution is calculated for an exact
local synchrotron spectral emissivity while analytically calculating 
the electron energy distribution everywhere by following its evolution
from the shock front (where it is assumed to be a pure power-law) due
to radiative and and adiabatic cooling. The light curve computed by using
a simplified emission model (equation \ref{eq:nocool}) that neglects electron cooling
altogether is an very good agreement with the GS02 semi-analytical results
(see Figure \ref{fig11}). The light curve computed by using an approximated
electron cooling presents three breaks
at low frequencies (corresponding to the transitions 
$\nu < \nu_c < \nu_m$ with the scaling $F_\nu \propto t^{1/6}$ $\rightarrow$
$\nu < \nu_m < \nu_c$ with $F_\nu \propto t^{1/2}$ $\rightarrow$ 
$\nu_m < \nu < \nu_c$ with $F_\nu \propto t^{3 (1-p)/4}$ $\rightarrow$
$\nu_m < \nu_c < \nu$ with $F_\nu \propto t^{(2-3p)/4}$)
and two breaks at high frequencies (corresponding to 
$\nu < \nu_c < \nu_m$ with $F_\nu \propto t^{1/6}$ $\rightarrow$
$\nu_c < \nu < \nu_m$ with $F_\nu \propto t^{-1/4}$ $\rightarrow$ 
$\nu_c < \nu_m < \nu$ with $F_\nu \propto t^{(2-3p)/4}$).
As can be noticed in Figure \ref{fig11}, our estimation of the cooling break 
frequency $\nu_c$ assuming that the electrons cool at their current local 
cooling rate over the dynamical time (see equations \ref{eq:nupl})
underestimate the cooling frequency determined by GS02.
For instance, an increase in $\nu_c$ of a factor of 4
produces a better agreement with the GS02 results (Figure \ref{fig11}, right panel).
It is worthwhile to stress that, while the mapped BMK light curve
and the GS02 results are applicable only for (highly) relativistic flows, 
the light curve computed from the numerical simulations is valid during 
all the deceleration of the flow to non-relativistic speeds.
Finally, we notice that at $\nu \lesssim 10^9$~Hz, self-absorption dominates 
and the light curves computed with our simple prescription are inaccurate.


\section{Numerical Tests}

We present in this section a series of one-dimensional
shock tubes and multi-dimensional tests.

\subsection{One-dimensional shock tubes}

Shock tube tests are used as 
standard tests as they are simple to implement and the exact analytical solution 
is known.
The tests were performed using a grid with size $0\le x \le1$, 
with an initial discontinuity at $x=0.5$. Here and in the following, we
refer to the left/right hand side of the discontinuity with the suffixes $_{L/R}$. 
In all the tests, we use a grid with 50 cells at the lowest level, with 4 levels
of refinement, corresponding to an effective resolution of 400 cells. We also make high 
resolution runs of the same tests, employing 400 cells at the lowest level, with 4 levels
of refinement, corresponding to an effective resolution of 3200 cells.
The Courant number if fixed equal to 0.8 in all tests, with a final integration time 
of $t=0.4$. The politropic index is fixed equal to 4/3 in the first shock tube test
and 5/3 in all others tests. As described in the following, in all the tests the exact 
solution is properly recovered.

The first test consist of a low-relativistic flow with a left state given by
$p_L=1$, $\rho_L=1$, $v_L=0.9$, corresponding to a Lorentz factor
of $\Gamma \approx 2.3$, and a right state given by $p_R=10$, $\rho_R=1$, $v_R=0$.
The evolution of this shock tube consists of two shocks and a stationary
contact discontinuity. Small oscillations, similar to those observed by
previous authors \citep[e.g.][]{2004A&A...428..703L,Wang08}, are present in the post-shock region.

The second shock tube consists of a low-relativistic flow 
with a left state given by $p_L=10$, $\rho_L=1$, $v_L=-0.6$ 
and a right state given by $p_R=20$, $\rho_R=10$, $v_R=0.5$.
In this test, two rarefaction wave are produced, together with
a left moving contact discontinuity. Both rarefaction waves
are properly recovered, while the contact discontinuity is smeared 
over $\sim 10$ cells.

\begin{figure}
\centering
 \includegraphics[width=0.45\textwidth]{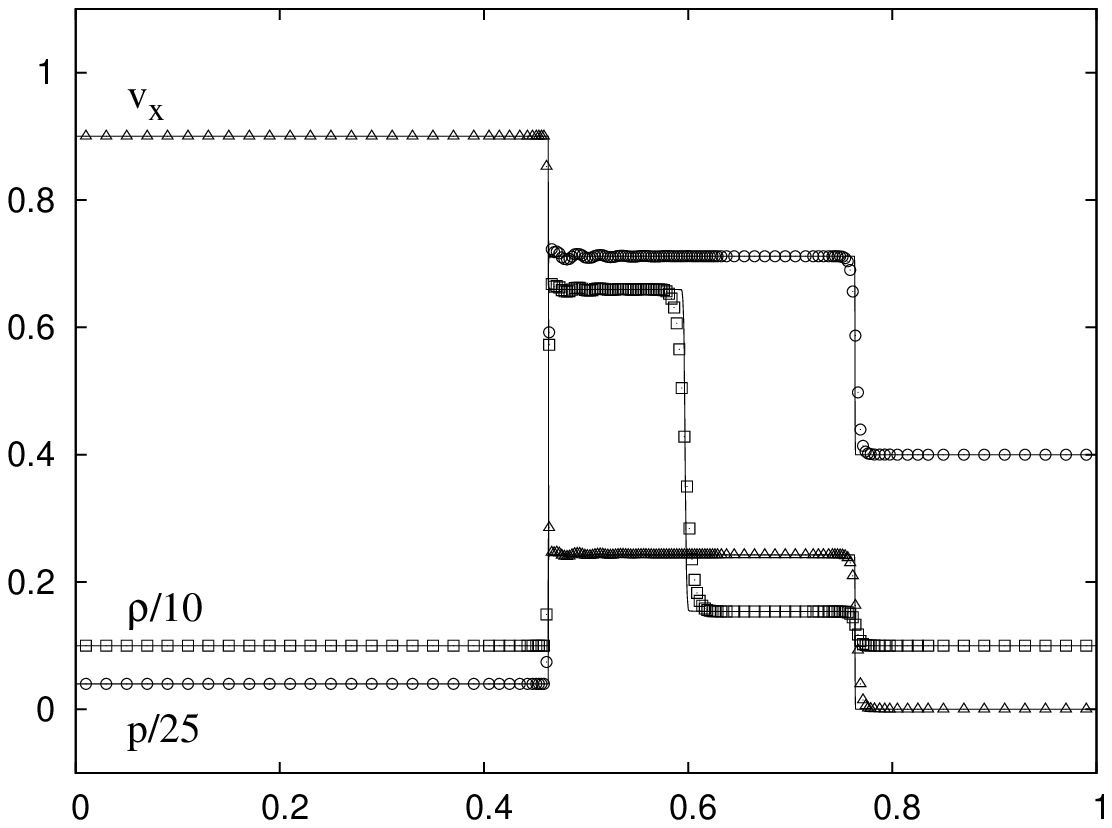}
 \includegraphics[width=0.45\textwidth]{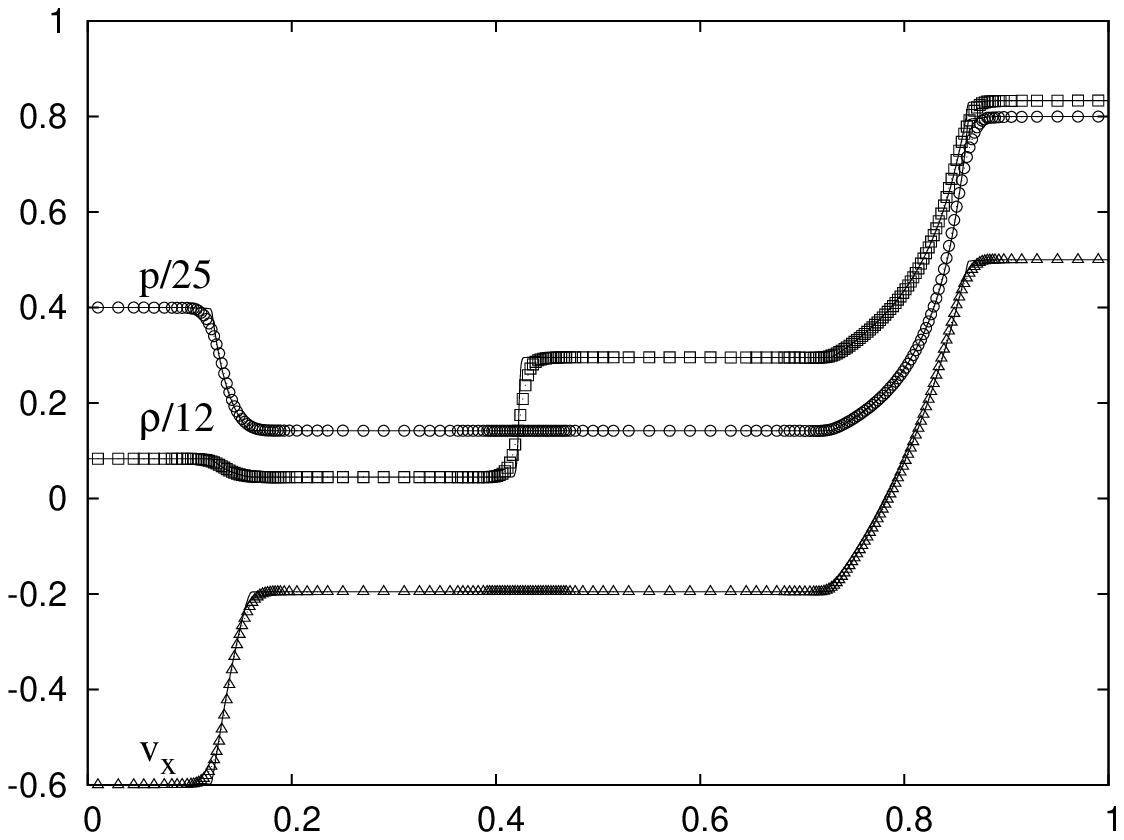}
 \includegraphics[width=0.45\textwidth]{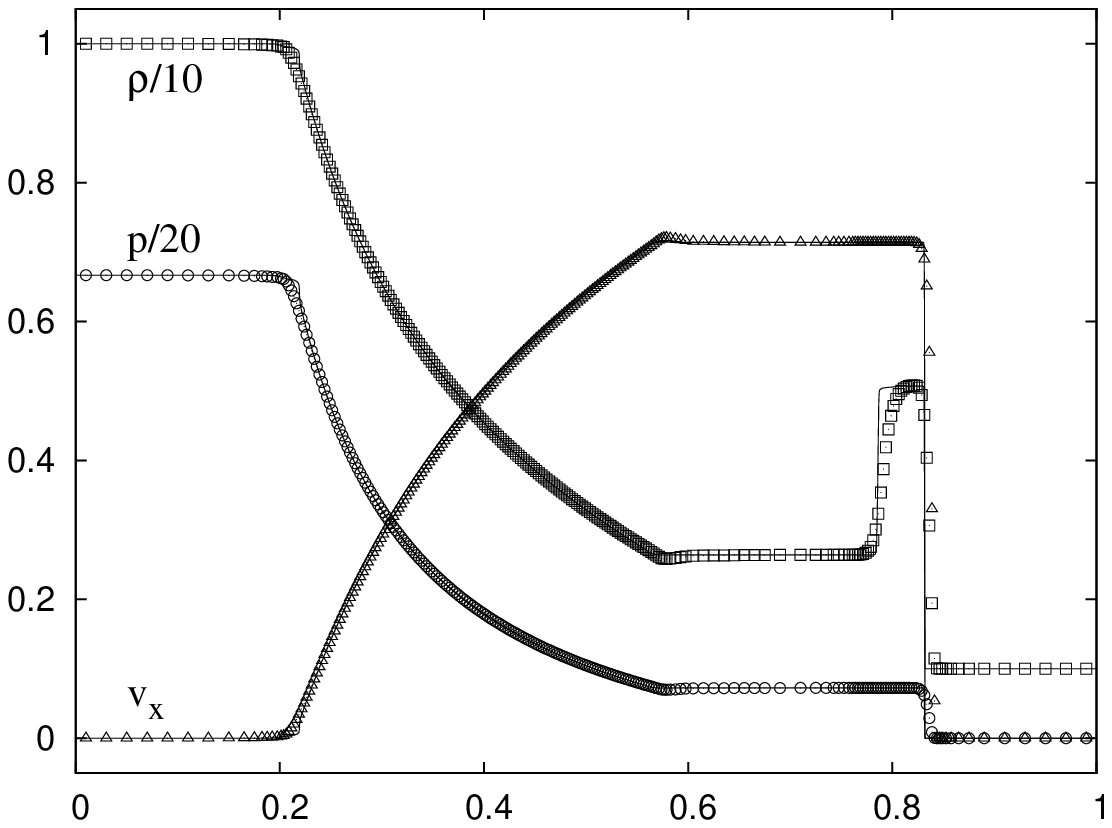}
 \includegraphics[width=0.45\textwidth]{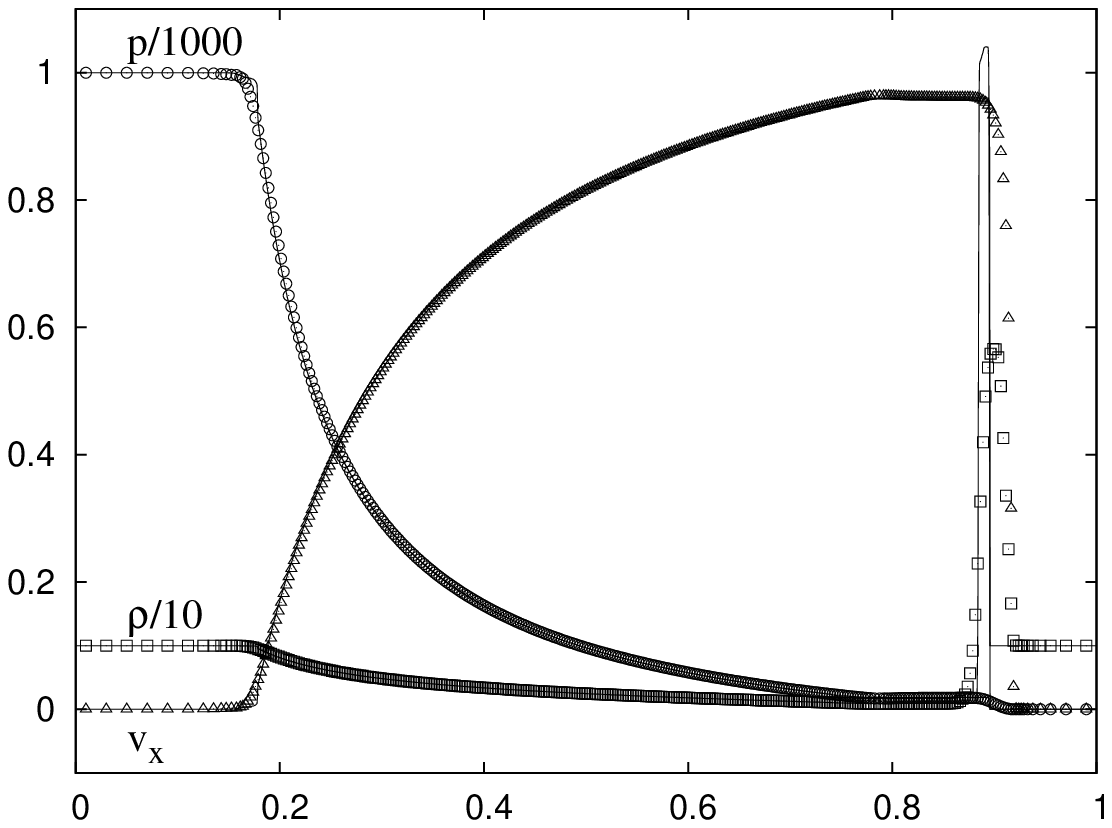}
\caption{One-dimensional shock tube problems at $t=0.4$. 
The variables shown are: density, velocity and pressure. 
The initial discontinuity was set at $x=0.5$, the Courant number is 
equal to 0.8, with a maximum resolution of 400 cells (points)
and 3200 cells (lines). The panels correspond to first (upper left),
second (upper right), third (bottom left) and fourth (bottom right)
shock tube tests (see the text for a detailed description of the 
initial conditions).}
\label{fig:testst}
\vskip 0.5cm
\end{figure}

The last two tests are taken from \citet{1998JCoPh.146...58D}, and refer to
blast wave explosions.
The third shock tube consists of  of a left state given with
$p_L=40/3$, $\rho_L=10$, and a right state given by $p_R=10^{-6}$, 
$\rho_R=1$, while in the last test the left state is given by 
$p_L=1000$, $\rho_L=1$, and the a right state is given by 
$p_R=0.01$, $\rho_R=1$, 
The large pressure gradient produces a mildly relativistic
shock (test 3) and a highly relativistic shock (test 4) 
with $\Gamma \approx 6$. As can be seen in Figure~\ref{fig:testst},
the solution consists in both cases of a strong shock moving
to the right and a rarefaction wave moving to the left. 
No oscillations are present in the solution. The shock is 
resolved within $\sim 4$ cells, while the contact discontinuity
is smeared over several cells. That is expected, due to the intrinsic
diffusive properties of the HLL schemes. In the second blast wave problem,
the size of the thin dense shell in the post-shock region
consists of only $\approx 4$ cells with the resolution employed.
As a consequence, the exact value of the density is not recovered 
at low resolution. However, this region is properly resolved in the high 
resolution run.


\subsection{Multi-dimensional tests}

\subsubsection{Relativistic 2D Riemann problem}

This test has been studied in the non-relativistic case by
\citet{LL98}, and extended to the SRHD case
by \citet{2002A&A...390.1177D}. It has been widely used recently
as a test for multi-dimensional SRHD codes
\citep[e.g.][]{2004A&A...428..703L,Wang08}.
The computational domain (at $t=0$) is divided in four regions:
\begin{eqnarray}
(\rho, v_x, v_y, p)^{NE} = (0.1,0,0,0.01)\qquad \mathrm{if}\qquad x \ge 0.5, y \ge 0.5 \nonumber \\
(\rho, v_x, v_y, p)^{NW} = (0.1,0.99,0,1)\qquad \mathrm{if}\qquad x \le 0.5, y \ge 0.5 \nonumber \\
(\rho, v_x, v_y, p)^{SW} = (0.5,0,0,1)   \qquad \mathrm{if}\qquad x \ge 0.5, y \le 0.5 \nonumber \\
(\rho, v_x, v_y, p)^{SE} = (0.1,0,0.99,1)\qquad \mathrm{if}\qquad x \le 0.5, y \le 0.5 \nonumber
\end{eqnarray}

We use a uniform grid with $400 \times 400$ cells, an adiabatic equation
of state with constant $\gamma=5/3$, and outflows boundary conditions.
The simulations ends at $t=4$. 
To better resolve the contact discontinuity, a more compressive MC limiter
is used here.
The results are shown in Figure~\ref{fig:test6}. The initial 
discontinuities across the four regions of the grid produce 
stationary contact discontinuity (with jumps in transverse velocities) 
between SW-NW and SE-SW, and shocks between NE-NW and SE-SW. These shocks
produce an elongated jet-like structure on the diagonal. These features, together
with the curved shock in the SW region, are qualitatively similar to those
obtained by previous authors. 


\begin{figure}
\centering
 \includegraphics[width=0.5\textwidth]{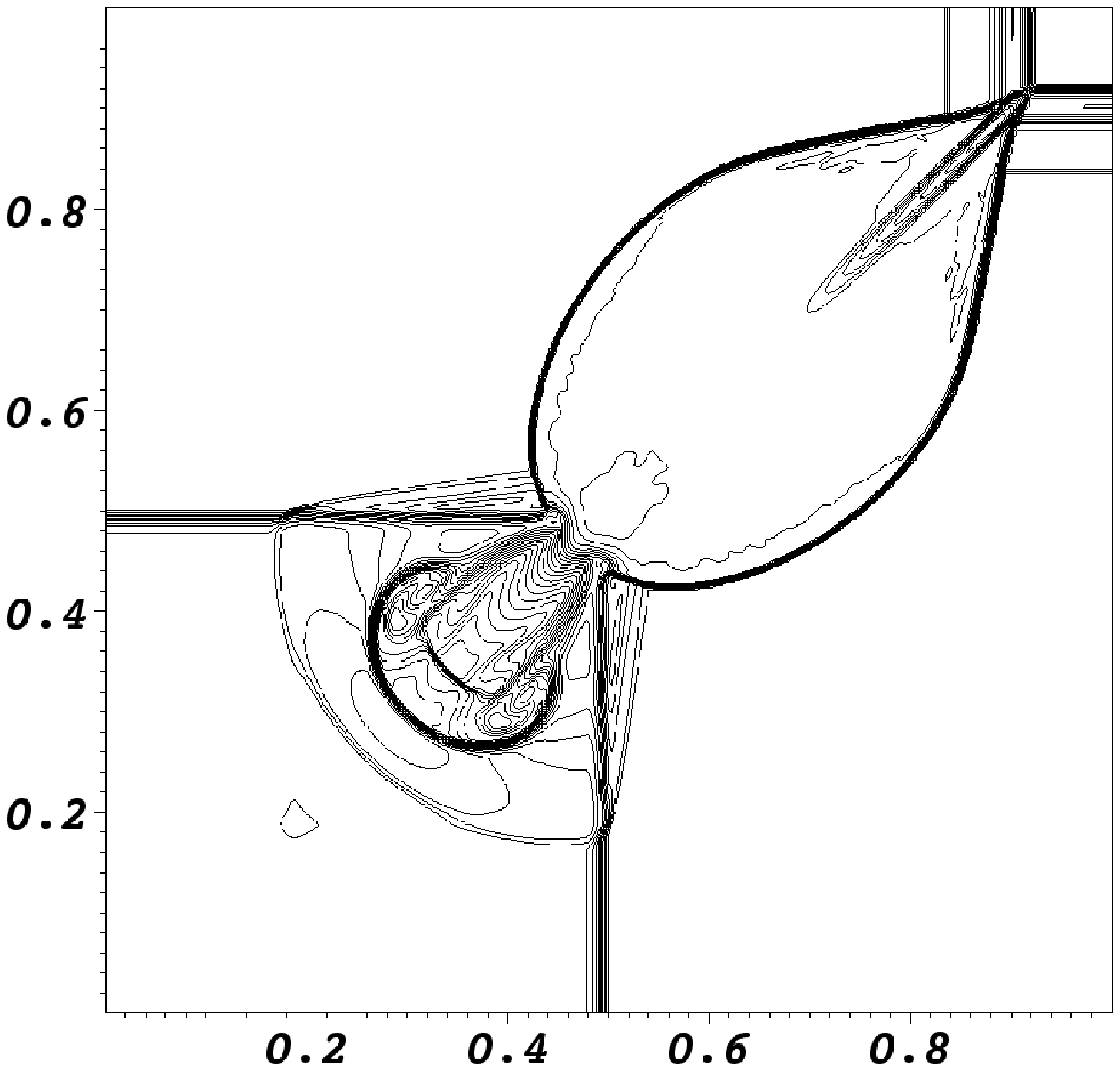}
\caption{Logarithm of the density for the Relativistic 2D
Riemann problem at $t=0.4$. Thirty equally spaced contours 
are plotted in the Figure.
}
\label{fig:test5}
\vskip 0.5cm
\end{figure}


\begin{figure}
\centering
 \includegraphics[width=0.5\textwidth]{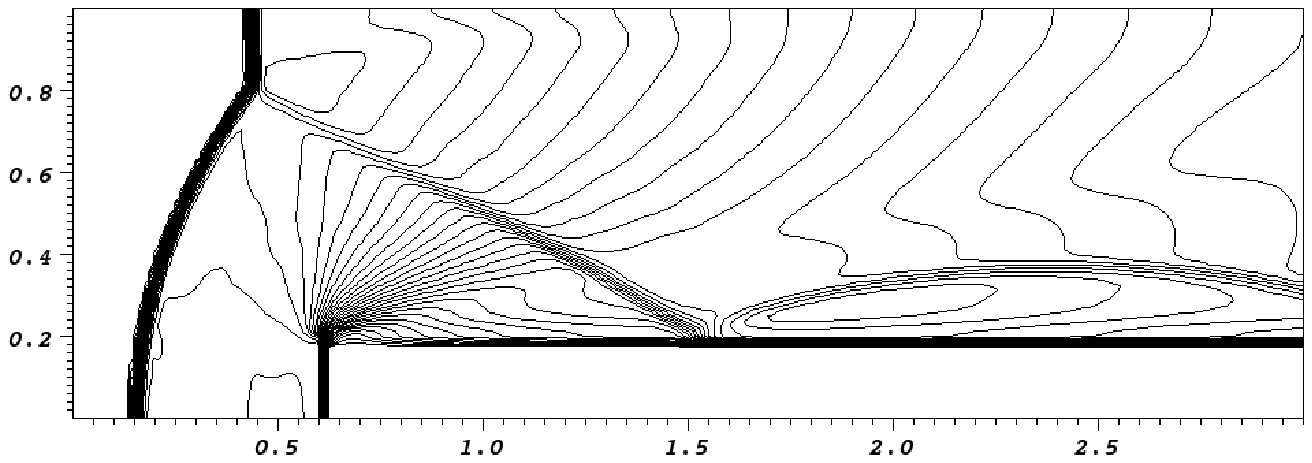}
\caption{Logarithm of the density for the Emery step problem
at $t=4.26$. Thirty equally spaced contours are plotted 
in the Figure.
}
\label{fig:test6}
\vskip 0.5cm
\end{figure}

\subsubsection{Emery step}

The ``Emery step'' test has become a standard test both
for non-relativistic and relativistic
hydrodynamics codes, and it consists of a wind moving through a tunnel.
Our initial conditions closely follow those by \citet{2004A&A...428..703L}.
 A relativistic flow moves initially horizontally
with velocity $v_x = 0.999 c$, corresponding to a Lorentz factor
of $\Gamma\approx 7$. The density is initially fixed at $\rho=1.4$
everywhere, with a pressure of $p=1/9$ and an adiabatic index 
of $\gamma=7/4$, corresponding to a Newtonian Mach number of
$M=3$.
The size of the tunnel is $0 \le x \le 3$ and $0 \le y \le 1$. 
A step is located in the region defined by $x \ge 0.6, y \le 0.2$. 
Inflow boundary conditions (with the same values used to fill the
tunnel initially) are fixed at the left boundary. Outflow boundary 
conditions are fixed at the right boundary, while reflecting boundary
conditions are fixed at the upper, lower and step boundaries. We use 
a uniform grid with $240 \times 80$ cells, with the HLL method coupled 
to the MC limiter.

Figure~\ref{fig:test6} shows the density stratification at $t=4.26$.
As the relativistic flow collides with the step, a reverse shock is formed.
This shock front is reflected from the upper boundary forming a 
stationary Mach stem. The results of this tests are similar to those of
\citet{2004A&A...428..703L}.


\end{document}